\documentclass[aps,prx,10pt,superscriptaddress,twocolumn,preprintnumbers,amsmath,amssymb]{revtex4-2}
\usepackage{hyperref}
\hypersetup{colorlinks=true,linkcolor=blue}
\usepackage{graphicx}
\usepackage{mathtools}
\usepackage{slashed}
\usepackage[caption=false]{subfig}
\usepackage{color}
\usepackage{xcolor}
\usepackage{bm}
\usepackage{booktabs}
\usepackage{float}

\newcommand{\e}{\mathrm{e}}

\begin{document}

\title{Induced Lorentzian geometry from the fermionic quantum metric}

\author{Patrick Copinger}
\email{copinger@hiroshima-u.ac.jp}
\affiliation{International Institute for Sustainability with Knotted Chiral Meta Matter (WPI-SKCM${}^2$), Hiroshima University, 1-3-1 Kagamiyama, Higashi-Hiroshima, Hiroshima 739-8531, Japan}
\author{Pablo Morales}
\email{pablo\_morales@araya.org}
\affiliation{Research Division, Araya Inc., Tokyo 101-0025, Japan}
\affiliation{Centre for Complexity Science, Imperial College London, London SW7 2AZ, UK}

\begin{abstract}
Quantum geometry is conventionally understood in parameter or momentum space and need not be related to a geometry in configuration space. Here we show for Dirac fermions in a background electromagnetic field that an induced Lorentzian geometry is generated from the fermionic quantum metric. We study a Lorentz covariant phase space worldline path integral construction, whose application of the adiabatic theorem onto spin degrees of freedom leads to the appearance of the Berry phase at $\mathcal{O}(\hbar)$, and the quantum metric at $\mathcal{O}(\hbar^2)$ in the worldline action. Both the Feynman and retarded/advanced propagators are examined in 3+1 and 2+1 dimensions, whose different on-shell implementations lead to related quantum metrics; QED/QED${}_3$ and condensed matter Dirac theory are considered. An effective theory is found by integrating out the momenta degrees of freedom, which gives rise to an emergent Lorentzian metric in configuration space, whose perturbation to the flat metric is dependent on the quantum metric and the background field strength. Apart from the spin degrees of freedom encoded into the Berry phase, the effective theory describes a scalar particle moving in the induced quantum geometry. Phenomenological applications to the covariant geometry are further studied using spectral holonomy in a gapped 3-dimensional Dirac material.
\end{abstract}
\maketitle

\section{Introduction}

The concept of geometry in physics is profound, impacting virtually every discipline. Not only is geometry present in e.g. the very makeup of spacetime, but is also present in the Hilbert space of wavefunctions--making up a quantum geometry. A prominent component quantum geometry is the Berry curvature~\cite{doi:10.1098/rspa.1987.0131}, which is associated with the geometric phase accumulated along a closed path in parameter space. Berry phases have famously described the Hall current and conductivity~\cite{Thouless:1982zz,*PhysRevLett.75.1348,*PhysRevB.53.7010}, the anomalous Hall effect~\cite{PhysRev.95.1154,*doi:10.1126/science.1058161,*PhysRevLett.88.207208,*doi:10.1126/science.1089408}, spin transport~\cite{doi:10.1126/science.1087128,*PhysRevLett.92.126603}, and even the chiral anomaly~\cite{Stephanov:2012ki,*Chen:2013iga,*Chen:2014cla}. The complementary real part of the quantum geometric tensor (QGT), $Q_{ab}=\mathfrak{g}_{ab}-(i/2)S_{ab}$, is the quantum metric (QM), $\mathfrak{g}_{ab}$, which is the pullback of the Fubini–Study metric on projective Hilbert space~\cite{PhysRevA.36.3479}.

On the manifold of quantum states the QM provides a natural Riemannian geometry that quantifies the distinguishability of neighboring states and thereby encodes information about the underlying local Hilbert space structure~\cite{Provost:1980nc,Anandan:1990fq}. The QM may be defined as the infinitesimal distance between neighboring quantum states for quantum parameter $\lambda$ such that $ds^2=\mathfrak{g}_{ab}d\lambda^a d\lambda^b$~\cite{Provost:1980nc,Kolodrubetz_2017}. In contrast to the Berry curvature, which characterizes a geometric phase that underlies a variety of topological phenomena, the QM measures the magnitude of the change of the quantum state under parameter variations thereby encoding local geometric information. The important distinction between the two is well-established in condensed matter, where the QM is present in a wide range of physical phenomena~\cite{yu2025quantumgeometryquantummaterials}, including Wannier-state localization~\cite{Resta_2011,*Marzari_2012}, orbital susceptibility~\cite{gao2015geometricaleffectsorbitalmagnetic,*Pi_chon_2016}, nonlinear optical responses~\cite{gao2019semiclassicaldynamicsnonlinearcharge,*Ahn_2021,*Ma_2021}, anomalous transport~\cite{Gao_2014,*Gao_2023}, exciton properties~\cite{srivastava2015signaturesblochbandgeometryexcitons}, and superfluid weight in flat bands \cite{Peotta:2015gmk}. Indeed, even when conventional band dispersion vanishes in superfluidity the QM exhibits a geometric contribution~\cite{Peotta:2015gmk,Herzog_Arbeitman_2022}. Also, the geometry of the QGT is directly experimentally accessible~\cite{b4fe5360382e4d3ca29564ec815322a6,*Yu:2019hsq}. While the QM is conventionally regarded as a geometry in parameter or momentum space, its interpretation as a physical metric raises a deeper question: can this quantum-state geometry itself be related to an emergent spacetime geometry?

In this manuscript we affirmatively probe this question with an analysis of the QGT of spin d.o.f. Since spacetime geometry puts time on an equal footing to space we are obliged to use a Lorentz covariant construction of the underlying fermionic quantum field theory (QFT). However, the Berry phase is readily analyzable in a first-quantized setting (such as appears for a single particle in the quantum mechanical path integral), in contrast to more familiar second-quantized settings for QFTs. Thus, we make use of the first-quantized worldline formalism~\cite{UsRep,ChrisRev} for a spinor, specifically the phase-space worldline formalism~\cite{Migdal:1986pz}, in which the avenue to the Berry phase and hence QGT are manifest~\cite{Copinger:2022jgg}.

The worldline formalism, originally developed by Feynman~\cite{Feynman:1950ir,*Feynman:1951gn}, then Strassler~\cite{Strass1}, wherein QFT observables may be represented as a path integral sum over relativistic trajectories has enjoyed research in strong field quantum electrodynamics (QED)~\cite{Reuter:1996zm, *Schmidt:1993rk, *Shaisultanov:1995tm, *Dunne:2002qf, *Dunne:2002qg, *Alexandrou:1998ia}, quantum chromodynamics (QCD) and non-Abelian theories~\cite{ChrisRev,103,*Hattori:2023egw,*Ahmadiniaz:2015xoa,*Mueller:2019gjj,*Copinger:2020feb}, and even axial gauge fields~\cite{Bastianelli:2024vkp,*Copinger:2022bwl}. Its usage is best known for the study of non-perturbative effects such as the worldline instanton techniques~\cite{Affleck:1981bma,*Dunne:2005sx,*Dunne:2006st} for studying the Schwinger effect~\cite{Schwinger:1951nm,*Dunne:2004nc,*Fecit:2025kqb}, or in the perturbative picture in which multiple Feynman diagrams may be incorporated on the worldline for constant electromagnetic field~\cite{Ahmad_2017}, plane wave~\cite{Edwards:2021vhg,*Edwards:2021uif,*Schubert:2023gsl,Copinger:2023ctz}, or even classical scattering of gravitons~\cite{Jakobsen:2022psy,*Dlapa:2022lmu,*Damgaard:2023vnx,*Jakobsen:2021smu} backgrounds to name a few. The worldline formalism has also been extended to in-in observables~\cite{fradkin1991quantum,*Copinger:2025ovz}, and may be treated with Monte-Carlo numerical techniques~\cite{Gies:2001tj,*Gies:2001zp}.

We construct an approach to the application of the adiabatic theorem onto the fermionic worldline phase-space action leading to a QM at $\mathcal{O}(\hbar^2)$, similar in spirit to the case of Weyl fermions in a quantum mechanical setting in~\cite{Mameda:2025rfn}. This is accomplished for numerous systems, namely in (2+1) and (3+1)-dimensions, and for both the Feynman and advanced/retarded propagators--it turns out their QMs are different, though with similar features. The Feynman QM to the best of our knowledge is new. We also analyze both using coordinates both for quantum electrodynamics (QED) and condensed matter, making our target scope interdisciplinary by construction. Using the QM in phase space we show how to arrive at an emergent metric in configuration space by integrating out the momenta d.o.f. The effective worldline Lagrangians we report are important in that we show how one may begin in flat space, and after employing the adiabatic theorem (on spin), we are left with a unique induced geometry that depends on the QM and electromagnetic field strength in (linearized to $\mathcal{O}(\hbar^2)$) curved space. 

The emergence of spacetime is a popular concept in physics. It is known to arise from non-commutative field theories for both complex scalars~\cite{RIVELLES2003191} and gauge field theories~\cite{doi:10.1142/S0217732306021682,*doi:10.1142/S0217751X0904587X}, and from matrix theories~\cite{Harold_Steinacker_2007,*STEINACKER20091}. Spacetime emergence has also been studied on the worldline for toy theories inspired by a Berry phase in~\cite{Copinger:2022srm}. Related notions have also been studied in condensed matter in ${^3}$He superfluids~\cite{Volovik:2003fe}, in Bose-Einstein condensates~\cite{Barcelo:2005fc}, and from corrugations, in the entanglement of many-body states~\cite{Swingle:2009bg} in graphene~\cite{Morales:2022zws}.

The manuscript is organized as follows: In Sec.~\ref{sec:setup} we outline the various physical setups examined, ending with a summary of our results. In Sec.~\ref{sec:phase_space} we discuss adiabaticity as it applies to the phase-space worldline formalism. And in Secs.~\ref{sec:Feynman} and~\ref{sec:retarded} we show an emergent geometry for respectively the Feynman and advanced/retarded propagators, first in momentum space and then as an induced Lorentzian metric in configuration space. In Sec.~\ref{sec:holonomy} we illustrate a phenomenological application using spectral holonomy. And finally in Sec.~\ref{sec:conclusions} we offer an outlook, along with future directions.

We use SI units throughout with $\hbar$ and $c$ or $v_F$ (where appropriate) present. Flat space indices are denoted with Roman letters, $a,b,\cdots$; and curved space indices with Greek letters. However, for spatial flat space indices we will exclusively use the letters ``$i,j,k$," or where appropriate bold font, e.g. $\bm{p}$. We use a mostly minus signature in flat space: $\eta^{ab}=\mathrm{diag}(1,-1,-1,-1)$ for (3+1)-dimensions or $\eta^{ab}=\mathrm{diag}(1,-1,-1)$ for (2+1)-dimensions. Our completely antisymmetric tensor,  $\varepsilon^{abcd}$ ($\varepsilon^{abc}$), is defined with $\varepsilon^{0123}=1$ ($\varepsilon^{012}=1$).  The covariant derivative is $D_{a}=\partial_{a}+i(q/\hbar)A_{a}$.  The covariant Weyl matrices in (3+1) are defined as $\sigma^a=(I_2,\boldsymbol{\sigma})$ and $\bar{\sigma}^a=(I_2,-\boldsymbol{\sigma})$.  Our Dirac matrices in (3+1) are given in the Weyl representation with
\begin{equation}
   \gamma^{a}=\begin{array}{cc}\begin{pmatrix}& \sigma^a\\\bar{\sigma}^a\end{pmatrix}\end{array},\;
   \gamma_{5}=\begin{array}{cc}\begin{pmatrix}-I_2\\& I_2\end{pmatrix}\end{array}\,.
\end{equation}
The spin tensor is defined as $\sigma_{ab}=(i/2)[\gamma_{a},\gamma_{b}]$. In (2+1)-dimensions we use for our Dirac matrices $\gamma^0=\sigma_z$, $\gamma^1=i\sigma_y$, and $\gamma^2=-i\sigma_x$. In both (3+1) and (2+1) we have that $\{\gamma_a,\gamma_b\}=2\eta_{ab}$.

\section{Setups and motivation}
\label{sec:setup}

The semi-classical limit of QFTs is a widely studied regime most notably offering the criteria for steepest descents in the path integral where one expands about classical solutions. Let us go ahead and characterize the criteria as $\hbar \ll 1$, or rather $\hbar/S \ll 1$ for action $S$, which from hereafter we assume is the dimensionless parameter governing the limit. When $S$ is a scalar weighted function, well-known semi-classical techniques may be readily employed. However, consider a matrix weighted action, for example appearing with path ordering such that under the path integral one has $\mathcal{P}\exp(iS/\hbar)$. A wide array of techniques is at one's disposal to treat such cases, including for example rewriting the path ordered exponential as a path integral over e.g., grassman variables~\cite{ChrisRev} or colored matrices~\cite{Barducci:1980xk,*Bastianelli:2013pta}. Additionally, the criteria $\hbar\ll 1$ also supports the adiabatic theorem leading to the geometric or Berry phase~\cite{doi:10.1098/rspa.1987.0131}, wherein a quantum phase is induced from a holonomic path in some parameter space of the Hamiltonian. Typically, for applications of the Berry phase onto a quantum mechanical system one uses the adiabatic theorem wherein a large time elapse is assumed, i.e. $\Delta t\gg 0$. However with a simple redefinition, the time elapse corresponds to the semi-classical limit. Let us demonstrate this. For a quantum mechanical system with Hamiltonian, $H$, one may take for its unitary evolution operator the path ordered exponential and integrated time redefinition the following:
\begin{equation}\label{eq:hbar_t_connection}
    \mathcal{P}\exp\Bigl\{ -\frac{ic}{\hbar}\int^{\Delta t}_0\!dt \,H(t) \Bigr\}\to
    \mathcal{P}\exp\Bigl\{ -\frac{ic\Delta t}{\hbar}\int^{1}_0\!dt' \,H(t') \Bigr\}\,,
\end{equation}
where one can clearly see that the semi-classical limit, $\hbar\ll 1$ corresponds to the adiabatic limit $\Delta t\gg1$ within the path ordered exponential. While the study of Berry phases and the QGT tensor are indeed well known for several non-relativistic setups in quantum mechanics and condensed matter, little is known about QFTs including QED, serving as a primary motivation of this work. However, our interest is not confined to just QED, but in fact similar setups are also present in condensed matter, which we approach with an equal motivation.

Our target environments include both high energy--(3+1)-dimensional quantum electrodynamics (QED) and the (2+1)-dimensional QED${}_3$--and the habitat where Berry phases most conventionally appear in condensed matter. In condensed matter we consider gapped Dirac semimetals, gapped graphene and Chern insulators. Let us first consider the free (3+1) and (2+1)-dimensional condensed matter cases with respective band Hamiltonians
\begin{align}\label{eq:H31}
    H_{\bm k}&=\hbar v_F \bm{k} \cdot\bm{\alpha} +\Delta\beta,\\
    H_{\bm k}&=\hbar v_F 
    (k^1\sigma^1 +k^2\sigma^2)
    +\Delta\sigma^3\,.
    \label{eq:H_2+1}
\end{align}
The gap $\Delta$ plays the role of a rest energy, while the Fermi velocity, $v_F$, replaces the speed of light, $c$.  The corresponding rest momentum is $\Delta/v_F$, and the positive energy corresponding to either Hamiltonian above is 
\begin{equation}
    E_{\bm k}=\sqrt{\hbar^2 v_F^2{\bm k}^2+\Delta^2}\,.
\end{equation}
One also has that ${\bm p}=\hbar {\bm k}$ for momentum $\bm{p}$. Likewise, we may equally well study the systems of QED and QED${}_3$, with only a simple substitution of units. Namely, with
\begin{equation}\label{eq:unit_map}
    mc\leftrightarrow v_F^{-1}\Delta\,,\quad \text{and}\quad
    c\leftrightarrow v_F\,.
\end{equation}
In this way we cover equivalently both condensed matter and high-energy setups. We will ultimately be interested in the more general setup in which a background field may be treated; in this case the minimally coupled Hamiltonian in 3+1 is
\begin{equation}\label{eq:H_full}
    H=v_FqA_{0}-v_F\bm{\alpha}\cdot(i\hbar\bm{\nabla}+q\bm{A})+\Delta\beta\,.
\end{equation}
And, the 2nd-quantized Lagrangian associated with the above is
\begin{equation}
    \mathcal{L}=\bar{\psi}(x)\bigl[i\hbar\slashed{D}-v_F^{-1}\Delta \bigr]\psi(x)\,,
\end{equation}
with covariant derivative $\slashed{D}_a=\partial_a+i(q/\hbar)A_a$. An equivalent Lagrangian in QED units is of course
\begin{equation}\label{eq:LQED}
    \mathcal{L}
    =\bar{\psi}(x)\bigl[i\hbar\slashed{D}-mc\bigr]\psi(x)\,.
\end{equation}

The condensed matter Hamiltonians above encompass a broad class of systems whose low-energy quasiparticles are described by massive Dirac fermions. In (2+1)-dimensions, this includes gapped graphene near an individual $K$ or $K'$ valley, where a sublattice-asymmetric potential generates the mass gap, as well as the continuum limit of Chern-insulator models and, within a fixed spin and valley sector, monolayer transition-metal dichalcogenides such as $\mathrm{MoS}_{2}$~\cite{Semenoff1984,*Haldane1988,*Xiao2012}. In (3+1)-dimensions, closely related massive Dirac Hamiltonians occur in the $\bm{k}\cdot\bm{p}$ description of narrow-gap and strongly spin-orbit-coupled materials, including the $\mathrm{Bi}_{2}\mathrm{Se}_{3}$ family of 3-dimensional topological insulators, and may likewise describe gapped Dirac-semimetal phases obtained after breaking the symmetry protecting a Dirac node~\cite{Zhang2009,*Liu2010}. The form in~\eqref{eq:H_full} applies when the electromagnetic background varies slowly on the lattice and interband length scales, so that a single Dirac sector may be treated locally and the field enters through minimal coupling. Correspondingly, momenta are restricted to the approximately linear portion of the band dispersion about the relevant Dirac point, while lattice-scale corrections, band anisotropy and curvature, intervalley scattering, additional remote bands, disorder, and strong interaction effects are neglected. In addition, throughout our subsequent adiabatic treatment we require the field-induced evolution to be sufficiently slow compared with the interband gap, and the induced momentum displacement sufficiently weak, that transitions between the separated positive- and negative-energy sectors remain suppressed; approaching gap closing or the strong-field Landau--Zener regime therefore lies outside the perturbative description employed below.

As part of this systematic approach to adiabaticity and the realization of the QGT in their QFT and condensed matter field theory settings, we analyze a variety of propagators in some detail. Specifically we will consider in this work the advanced/retarded and Feynman propagators:
\begin{align}\label{eq:Feyn_prop1}
    S^c(x,y) &= i\langle \Omega|\mathcal{T}\psi(x)\bar{\psi}(y)|\Omega\rangle\,,\\
    S^\textrm{ret}(x,y) &= i\theta(z_0)\{\psi(x),\bar{\psi}(y)\}\,,\\
    S^\textrm{adv}(x,y) &= -i\theta(-z_0)\{\psi(x),\bar{\psi}(y)\}\,,\label{eq:Feyn_prop3}
\end{align}
where $z_a\coloneqq x_a-y_a$ and $\mathcal{T}$ represents the causal ordering of the propagators in the usual sense. $\psi(x)$ is the fermion operator for either (2+1) or (3+1)-dimensional Hamiltonian. For simplicity we have confined our attention to systems in which the asymptotic vacuum states coincide, i.e., $|\Omega\rangle\sim|\textrm{in}\rangle\sim|\textrm{out}\rangle$, such that $|\langle \textrm{out}|\textrm{in}\rangle|^2\approx 1$, and we can ignore vacuum non-persistence. For a more complete picture in a background field see~\cite{Fradkin:1981sc}. In the advanced/retarded propagators $i\langle \Omega|\{\psi(x),\bar{\psi}(y)\}|\Omega\rangle=i\{\psi(x),\bar{\psi}(y)\}$ represents the anti-commutation function, or rather the evolution operator in the quantum mechanics of Dirac fermions. As it turns out adiabaticity impacts each propagator, whether Feynman or advanced/retarded, in a completely different way leading to different but related QGT quantities. We can take $A_a$ in our systems to represent either a background field or a fluctuating field to be integrated over, however, we will turn our attention to the case of a background field for concrete applications.

A general motivation for studying the various propagators is that they provide a natural setting in which to study the QGT and its QM, for application of the adiabatic theorem onto the spin d.o.f. within the path integral is well-suited.  Further, in the semi-classical limit the action of the propagators dictates the classical path. More specifically for each individual propagator there are unique advantages. For the Feynman propagator the semi-classical limit or adiabaticity naturally appears for scattering amplitude in which $t_\textrm{in}\to-
\infty$ or $t_\textrm{out}\to\infty$ asymptotic times are utilized. For the advanced/retarded propagators adiabaticity points to response, dissipation, and causal evolution. Before treating technical aspects of adiabaticity onto each propagator using the phase-space worldline formalism, let us spell out what we get from doing so.

\begin{table*}[t]
\begin{tabular}{c|c|c|c|c|c}
\toprule
Propagator & Dimension & Broken Group & $h_{\mu\nu}$ & $\mathcal{B}_\mu $ & s \\
\midrule
$S^c$&3+1 & Spin(3,1)/SO(3) & $\frac{q^2}{4m^{4}c^4}F_{\mu\rho}\bigl(\eta^{\rho\sigma}-\hat{p}^{\rho}\hat{p}^{\sigma}\bigr)F_{\sigma \nu}$ & $\frac{-q}{2(|p_{0}|+p)}\varepsilon^{0\sigma\nu\rho}\Sigma_{\rho}\hat{p}_{\nu}F_{\sigma \mu}$& $\frac{\slashed{p}\gamma_0+\epsilon p}{\sqrt{2p(p+|p_0|)}}$\\
$S^c$&3+1 & Cl${}^{\textrm{odd}}$(3,1) & $\frac{q^2}{4m^{4}c^4}F_{\mu\rho}\bigl(\eta^{\rho\sigma}-\hat{p}^{\rho}\hat{p}^{\sigma}\bigr)F_{\sigma \nu}$ & $\frac{q}{2p^{2}}\sigma^{\sigma\nu}p_{\nu}F_{\sigma\mu}$ &$\frac{1}{\sqrt{2}}(1-\gamma_{5}\slashed{\hat{p}})$\\
$S^{\textrm{ret}}$ &3+1 & Spin(3,1)/SO(3) & $\frac{q^2}{4p^{4}_0}F_{\mu i}\bigl(\eta^{ij}+\tfrac{1}{p_0^2}p^{i}p^{j}\bigr)F_{j \nu}$ &$\frac{-q}{2p_0(p_0+p)}\varepsilon^{0\sigma\nu\rho}\Sigma_\rho p_\nu F_{\sigma \mu}$ & $\frac{p_0+p+\gamma^i p_i}{\sqrt{2p_0(p_0+p)}}$\\
$S^c$&2+1 & Spin(2,1)/SO(2) & $\frac{q^2}{4m^{4}c^4}F_{\mu\rho}\bigl(\eta^{\rho\sigma}-\hat{p}^{\rho}\hat{p}^{\sigma}\bigr)F_{\sigma \nu}$ & $\frac{\epsilon q\varepsilon^{0\sigma\nu}}{2p(|p_{0}|+p)}p_{\nu}F_{\sigma\mu}\gamma^{0}$ &$\frac{\slashed{p}\gamma_0+\epsilon p}{\sqrt{2p(p+|p_0|)}}$ \\
$S^{\textrm{ret}}$ &2+1 & Spin(2,1)/SO(2) &$\frac{q^2}{4p^{4}_0}F_{\mu i}\bigl(\eta^{ij}+\tfrac{1}{p_0^2}p^{i}p^{j}\bigr)F_{j \nu}$  &$\frac{-rq}{2p_0(p_0+p)}\varepsilon^{0\sigma \nu}p_{\nu}F_{\sigma\mu}$  &$\frac{p_0+p+\gamma^i p_i}{\sqrt{2p_0(p_0+p)}}$ \\
\bottomrule
\end{tabular}
\caption{Summary of emergent spacetime results in SI units for QED; for conversion to condensed matter units see~\eqref{eq:unit_map}. Perturbations to the induced metric, $h_{\mu\nu}$, the Berry phase, $\mathcal{B}_\mu$, and similarity transform, $s$, as they appear in the effective single particle worldline action are shown for each propagator type, dimension, and corresponding broken symmetry after application of the adiabatic theorem. Note for the (3+1)-dimensional Feynman case one may additionally diagonalize the effective worldline Hamiltonian proportional to $\gamma_5$, thus resulting in the broken group Cl${}^\textrm{odd}(3,1)$. For reference here $p=\sqrt{p_0^2-\boldsymbol{p}^2}>0$ and $\hat{p}_\mu=p_\mu/p$.}
\label{tab:results}
\end{table*}

\subsection{Summary of results}

Application of the adiabatic theorem onto the spin d.o.f. within the Feynman and advanced/retarded propagators listed in eqs.~\eqref{eq:Feyn_prop1}-\eqref{eq:Feyn_prop3} leads to the appearance of a Berry phase at $\mathcal{O}(\hbar)$ and the QM at $\mathcal{O}(\hbar^2)$. The QM metric we find is expressed in momentum space, however upon integrating out the momenta we find an emergent Lorentzian spacetime geometry. Hence the various propagators given in eqs.~\eqref{eq:Feyn_prop1}-\eqref{eq:Feyn_prop3} can be expressed in a dual form, from fermionic d.o.f. to bosonic d.o.f. at the single particle Lagrangian level but in curved space, whose metric is governed by the QM and electromagnetic field strength. However, adiabaticity affects each propagator in a subtlety different way, largely due to the fact that the adiabaticity is applied before (after) momenta are taken on-shell, i.e., $p^2=m^2c^2$ in the Feynman (advanced/retarded) propagators. Here let us state the emergent spacetime dual representation for each propagator. For comparison purposes let us present the summary all in QED units; using~\eqref{eq:unit_map} one may see their equivalent forms in condensed matter units. Further the conventions differ slightly here with those given elsewhere in the text for brevity.

Beginning with the Feynman propagator we ultimately find for the single-particle worldline Lagrangian to $\mathcal{O}(\hbar^2)$
\begin{equation}\label{eq:L_eff1}
    \mathcal{L}_\textrm{eff} = -mc\sqrt{\dot{x}^\mu g_{\mu\nu}^\textrm{ind}(x)\dot{x}^\nu}-qA_\mu\dot{x}^\mu-\hbar\mathcal{B}_\mu\dot{x}^\mu\,.
\end{equation}
Here $g^\textrm{ind}_{\mu\nu}=\eta_{\mu\nu}+\hbar^2h_{\mu\nu}$ is the induced effective metric with perturbation to flat spacetime represented with $h_{\mu\nu}$, which depends (for any propagator or dimension) largely on the QM and the background field strength; we generally find that $h_{\mu\nu}\propto F_{\mu\alpha}\mathfrak{g}^{\alpha\beta}F_{\beta\nu}$. This provides an important example where the QM directly can produce an emergent spacetime, not just in the parameter space of Hamiltonian but in configuration space showing a unique spacetime induced geometry from the QM. $\mathcal{B}_\mu=qB^\nu F_{\nu\mu}$ contains the (non-)Abelian Berry phase, $B_\mu$. And $A_\mu$ and $F_{\mu\nu}$ are respectively the gauge field and strength. See Table~\ref{tab:results} for expressions of $h_{\mu\nu}$ and $\mathcal{B}_\mu$ given for each propagator for a given dimension and diagonalized element. (For the (3+1)-dimensional Feynman case, in addition to diagonalizing to $\gamma_0$ one may diagonalize to $\gamma_5$.) Momenta in the induced metric and in the Berry phase are determined from their classical paths obeying the Lorentz force e.o.m. or the velocity-momentum relation. The effective Lagrangian appears in the Feynman propagator through
\begin{equation}
S^c\simeq \frac{i}{2}(1+\slashed{\hat{p}}_x)s_x\int^x_y\!\mathcal{D}x\mu\sqrt{|g^{\rm ind}|}\,\mathcal{P}\e^{\frac{i}{\hbar}\int^1_0d\tau\mathcal{L}_{\textrm{eff}}} s_y^{-1}\,.
\end{equation}
Here $\mu\simeq \det (mc/\sqrt{\dot{x}^2})^{d/2}$ with $d$ representing the spatial dimension, and $\mathcal{P}$ denotes a path-ordering in the Dirac matrices.

Next the advanced/retarded propagator takes on a similar form. The induced geometry we find differs, but is similar, to the Feynman case. We begin as before with the worldline action; this takes the same form as it did for the Feynman case in~\eqref{eq:L_eff1}, however with the kinetic term $-mc \to -mcr$ for the retarded propagator, where $r$ can be positive or negative depending on which energy branch is being projected. The advanced propagator has $-mc\to mc r$. And the full propagators become
\begin{align}
    S^{\substack{\textrm{adv} \\ \text{ret}}}&\simeq \mp i\theta(\mp z_0)\sum_{r=\pm} P_x^r s_x\int^x_y\!\mathcal{D}x\mathcal{M}\sqrt{|g^{\rm ind}|}\,\notag\\
    &\quad\times\substack{\bar{\mathcal{P}} \\ \mathcal{P}}\e^{\frac{i}{\hbar}\int^1_0d\tau\,\mathcal{L}_{ \rm eff}^{(\mp r)}}s^{-1}_y\gamma_0\,,
\end{align}
where $\mathcal{P}$ ($\bar{\mathcal{P}}$) denote path ordering (anti-path ordering), $\mathcal{M}=\det{}[\Delta^{d/2}v_F^{-d/2}|\dot{x}_0 |(\dot{x}_0^{2}-\dot{\bm{x}}^{2})^{-(d+2)/4}]$, and the projection operators $\mathcal{P}^\pm$ projecting either positive or negative energy modes.

\section{The phase space worldline formalism and the adiabatic theorem}
\label{sec:phase_space}

\subsection{Feynman propagator}

Berry phases are well understood in the single particle picture of quantum mechanics, for example for a particle with spin undergoing cyclical motion in an inhomogeneous magnetic field with $H=\bm{\sigma}\cdot\bm{B}$, and it is intuitive to study the same kind of single particle picture for our fermionic QFT systems. This can be accomplished with the worldline formalism~\cite{Feynman:1950ir,*Feynman:1951gn,ChrisRev}, specifically the phase space worldline formalism~\cite{Migdal:1986pz}, wherein one expresses e.g. the QFT propagator in first-quantized form. Let us illustrate how to do this with the causal Feynman propagator first in~\eqref{eq:Feyn_prop1}. We begin from the path integral definition and then integrate out the fermions to produce an inverse operator governing the defining differential equation:
\begin{align}
    S^c(x,y)&=\int\mathcal{D}\psi\mathcal{D}\bar{\mathcal{\psi}}\,
    \psi(x)\bar{\psi}(y)\,\e^{\frac{i}{\hbar}\int d^4x\mathcal{L}}\,\\
    &=\langle x|\frac{-\hbar}{i\hbar\slashed{D}-mc+i0^+}|y\rangle\,,
\end{align}
where we have used~\eqref{eq:LQED}. Then introducing a Laplace transform identifiable with Schwinger propertime, $T$, we arrive at an integral representation with a spacetime kernel,
\begin{equation}
    S^c(x,y)=i\int^\infty_0dT\,\langle x|\e^{-\frac{i}{\hbar}[-i\hbar\slashed{D}+mc-i0^+]T}|y\rangle\,.
\end{equation}
The $i0^+$ factor ensures convergence in the IR at $T\to\infty$. Now the spacetime kernel may be expressed in path integral form through conventional manipulations~\cite{Copinger:2022jgg}. However, rather than use Schwinger propertime directly, we elect to use an einbein path integral. The final transformed path integral in coordinate space then reads
\begin{equation}
    S^c(x,y)=i
   \int_y^x \!\mathcal{D}x
   \int\frac{\mathcal{D}p}{2\pi\hbar}\mathcal{D}e\,\e^{\frac{i}{\hbar}S}\mathcal{W}\,,
   \label{eq:Green}
\end{equation}
with gauge field dependent part of the action being
\begin{equation}
    S\coloneqq\int_{0}^{1}d\tau[-(mc-i0^+)\,e-p_{a}\dot{x}^{a}
   -qA_{a}\dot{x}^{a}]\,,
   \label{eq:S_A}
\end{equation}
and the path ordered exponential with worldline Hamiltonian being
\begin{equation}
   \mathcal{W}\coloneqq\mathcal{P} 
   \exp\Bigl\{\frac{i}{\hbar}\int_{0}^{1}\!d\tau\,e\slashed{p}\Bigr\}\,.
   \label{eq:Wilson_Dirac}
\end{equation}
$p$ in the above represents the physical kinetic momentum of the system (by shifting $p\to p-qA$ one may arrive at the canonical momentum representation--this is valid due to the free boundary condition in the momentum path integral). The system is reparameterization invariant since one may equally well write $\tau\to f(\tau)$ and $x,p\to x(f),p(f)$ with $e\to (d\tau/df)e$. To arrive at the Schwinger propertime single integral one can insert a Faddeev-Popov gauge fixing functional, with appropriate ghost terms, to effectively send $e\to T$ and $\int \mathcal{D}e\to\int^\infty_0 dT$; see also~\cite{doi:10.1142/7305}. We have left implicit a volume factor in the einbein path integral which would be needed before gauge fixing; however, manipulations later on will fix the gauge automatically. However, due to the presence of $i0^+$ gauge choices are restricted such that $e(\tau)>0$ for convergence.

The distinct merit of using the phase space worldline construction is apparent in that in analogy to for example the Weyl Hamiltonian with $\bm{p}\cdot \bm{\sigma}$ we have a corresponding Lorentz invariant Hamiltonian, $\slashed{p}$, with which we can apply the adiabatic theorem leading to the Berry phase and properties of the QGT. In fact, all the non-trivial aspects are encoded in the path ordered exponential,~\eqref{eq:Wilson_Dirac}. The eigenvalues of $\slashed{p}$ are $\pm p$ with
\begin{equation}\label{eq:p}
    p\coloneqq \sqrt{p_0^2-\bm{p}^2}\,,
\end{equation}
and on-shell equal to $mc$ or $v_F^{-1}\Delta$ depending on usage. And the eigenvalues for $p_0^2>\bm{p}^2$ are real. This is another reason for using the phase space worldline representation rather than the conventional quadratic worldline representation~\cite{ChrisRev} in which momentum has already been integrated out. In the quadratic representation one has a path integral with path ordered spin factor exponential being $\exp((iq/\hbar)\int^1_0d\tau\, e \sigma_{ab}F^{ab})$ whose eigenvalues are complex and so too would be the ensuing Berry phase and QGT.

A motivation of our study lies in the semi-classical limit, and equally the adiabatic limit. However the adiabatic limit as it appears in~\eqref{eq:hbar_t_connection} is related to a real time and not propertime directly. Let us sketch how the two are in fact related in a distinctive property of the Feynman propagator, that is its connection to scattering amplitudes via LSZ reduction. The asymptotic in and out times, i.e. $t_{\rm in}\to -\infty$ and $t_{\rm out}\to \infty$, in scattering amplitudes have been shown in the worldline formalism for scalar and spinor matter lines in their quadratic forms to be related to the limit of large propertime, $|T|$, in~\cite{Bonocore:2020xuj,*Mogull:2020sak,Copinger:2023ctz}. For truncation on the outgoing state, the essential process extracts a $p^2-m^2c^2+i0^+$ factor from the path integral, then when truncating by multiplying by the same factor in the propagator, and taking the mass-shell condition, one can show the propertime integral reduces to a boundary integral projecting out the $T\to \infty$. The same process occurs in~\eqref{eq:Green}. Consider the Fourier transformed propagator corresponding to~\eqref{eq:Green}. Then one may extract the endpoint momentum via a shift $p(\tau)\to p(\tau)+p$, for endpoint momenta $p$; one then has a $\slashed{p}-mc+i0^+$ factor in the action. LSZ on the, e.g., outgoing state amounts to acting on the propagator with operator $\lim_{\slashed{p}\to mc-i0^+}[\slashed{p}-mc+i0^+]$, which projects the $T\to \infty$ limit of the path integral, equivalently $e\to T\to\infty$ in our setup. Thus the large asymptotic real time corresponds to a large propertime of the phase space worldline formalism in the LSZ reduction of Feynman propagator. We refer the reader to Refs.~\cite{Bonocore:2020xuj,*Mogull:2020sak,Copinger:2023ctz} for further details. 

Circling back to the semi-classical criteria of $\hbar\to 0$, let us emphasize here that paths in the Feynman phase space worldline path integral that are close to the on-shell condition, i.e., $p^2=m^2$, are dominant. This is readily apparent simply from the fact that we are looking at the semi-classical picture, however, let us digress. Consider the action composed of both~\eqref{eq:S_A}, and exponential argument of~\eqref{eq:Wilson_Dirac}; steepest descents in the einbein, $e$, enforces the linear matrix-weighted mass shell condition of $mc=\slashed{p}$ (in its quadratic form of course $m^2c^2=p^2$). Off-shell, or more notably paths for which $|p^2-m^2c^2|\gg 0$ are strongly suppressed. Thus, (also considering the fact that our background field does not break the vacuum-persistence criteria) we confine our attention in the Feynman propagator to paths in which
\begin{equation}\label{eq:p_constraint}
    p^2(\tau)>0\,,
\end{equation}
for either positive or negative $p_0$. This criterion furthermore, we will later show, gives a well-defined definition to the adiabatic theorem in the Feynman phase space worldline picture since the eigenvalues of $\slashed{p}$ are $\pm p$ with $p$ given by~\eqref{eq:p}, since off-shell $p$ becomes complex. Formally we may treat this by the restricted path integral $\int_\mathcal{C} \mathcal{D} p/(2\pi \hbar)^4$ with $\mathcal{C}=\{ p^a(\tau) | p^2(\tau)>0, p_0(\tau) \lessgtr 0  \forall \tau\}$. However, this is not a necessary reduction, as the semi-classical limit preferentially selects paths near the mass-shell. Finally, whether exactly on-shell or off-shell and with $p^2>0$ produces a different Berry phase and quantum metric. We will treat both with the on-shell version residing with the advanced and retarded propagators. 

Last, before treating the advanced and retarded propagators, in addition to treating both the (2+1) and (3+1)-dimensional Feynman propagators, in (3+1)-dimensions we also show two different ways to employ the adiabatic theorem, one in which we diagonalize $\slashed{p}$ to $\pm \mathrm{sgn}(p_0) p \gamma_0$ using a Spin$(3,1)$ similarity transform, and the other to $\pm \mathrm{sgn}(p_0) p \gamma_5$ using a Cl$(3,1)$ for the Clifford group. The advantage in diagonalizing to $\gamma_5$ we will show is that manifest Lorentz invariance is still present in the action, even with the \textit{Dirac matrix} d.o.f. In both $\gamma_0$ and $\gamma_5$ cases, however, a similar quantum metric and emergent spacetime is found.

\subsection{Advanced/retarded propagators}

Let us now formulate the closely related advanced/retarded propagators. One may immediately write down their $i0^+$ construction in operator form as
\begin{align}
    S^{\substack{\textrm{adv} \\ \text{ret}}}&=\langle x|\frac{-\hbar}{i\hbar\gamma_0\slashed{D}-mc\gamma_0\mp i0^+}|y\rangle\gamma_0\\
    &\to\langle x|\frac{-\hbar}{i\hbar\partial_{0}-v_F^{-1}H\mp i0^+}|y\rangle\gamma_0\,,
\end{align}
where the Hamiltonian is given by~\eqref{eq:H_full}. Even the above form, as it is expressed in (3+1)-dimensions, is unfamiliar. Therefore let us show its usual quadratic form in QED units for the free case with $A\to0$. There we may diagonalize the momenta to find the Fourier transformed representation as
\begin{align}
    S^{\substack{\textrm{adv} \\ \text{ret}}}&=\int \frac{d^4p}{(2\pi \hbar)^4}\frac{-\hbar\e^{-i\hbar^{-1}p\cdot z}}{\slashed{p}-mc\mp i0^+\gamma_0}\\
    &=\int \frac{d^4p}{(2\pi \hbar)^4}\frac{-\hbar(\slashed{p}+mc)\e^{-i\hbar^{-1}p\cdot z}}{p^2-m^2c^2\mp i\mathrm{sgn}(p_0)0^+}\,.
\end{align}
In the second line we have gone to the quadratic representation in the denominator; this has the familiar ${\rm sgn}(p_0)$ dependence in the $i0^+$ factor, demonstrating the validity of our setup.

One may equally well write the above kernels in Schwinger propertime form as
\begin{align}
    S^{\textrm{ret}}(x,y)&=i\int^\infty_0dT\,\langle x|\e^{-\frac{i}{\hbar}[-i\hbar\partial_{0}+v_F^{-1}H-i0^+]T}|y\rangle\gamma_0\,.\\
    S^\textrm{adv}(x,y)&=i\int_0^{-\infty}dT\,\langle x|\e^{-\frac{i}{\hbar}[-i\hbar\partial_{0}+v_F^{-1}H+i0^+]T}|y\rangle\gamma_0\,.
    \label{eq:Sadv}
\end{align}
and like before with the Feynman propagator we may construct a reparameterization invariant path integral 
\begin{equation}\label{eq:Sret_def}
    S^{\text{ret}}(x,y)= i
   \int_y^x \!\mathcal{D}x
   \int\frac{\mathcal{D}k}{2\pi}\mathcal{D}e\,\e^{\frac{i}{\hbar}S_{H}}\mathcal{W}_H\gamma_0\,,
\end{equation}
with action and path ordered exponential of the Hamiltonian
\begin{align}
    S_H&\coloneqq  \int^1_0d\tau\bigl[e\hbar k_0+i0^+ e-\hbar k_a\dot{x}^a-qA_a\dot{x}^a\bigr]\,,\\
    \mathcal{W}_H&\coloneqq \mathcal{P}\exp\Bigl\{ \frac{i}{\hbar}\int^1_0d\tau\,\bigl[  
    -v_F^{-1}\Delta\beta -\hbar\bm{k}\cdot\bm{\alpha}\bigr]e  \Bigr\}\,.
    \label{eq:W_H}
\end{align}
$S^{\text{adv}}$ is the same as above, however, with $0^+\to-0^+$ and with the path ordering in propertime going to an anti-path ordering with $\mathcal{P}\to\bar{\mathcal{P}}$. We take regardless of Dirac matrix representation in (3+1)-dimensions that $\beta\equiv \gamma_0$ and $\alpha^i \equiv \gamma_0 \gamma^i$; and that in (2+1) dimensions $\beta\equiv \sigma^3$, $\alpha^1\equiv \sigma^1$, and $\alpha^2=\sigma^2$. Also, for $S^{\text{adv}}$ one must have a gauge choice of $e(\tau)<0$ due to the $-i0^+$ factor. Noting the close similarity between the advanced and retarded propagators let us simply turn our attention to the retarded propagator noting that the advanced propagator can be had with simple substitutions.

We have stated that the advanced/retarded propagators are equivalent to the well-known quantum mechanical evolution operator and let us confirm this now. One may immediately integrate out $k_0$ in~\eqref{eq:Sret_def} to produce a functional delta, $\delta[e-\dot{x}^0]$ for $\dot{x}_0(\tau)>0$, which serves to fix the einbein gauge. The significance of this gauge then can be seen as a connection to the ordinary quantum mechanical path integral for 
\begin{equation}\label{eq:SH_qm}
    S_H\to \int^{x_0}_{y_0}dx_0\,\Bigl[ i0^+ +\hbar \bm{k}\cdot \bm{\dot{x}}-qv_F^{-1}\Phi+q\bm{A}\cdot\bm{\dot{x}}  \Bigr]
\end{equation}
where the `dots' now represent differentiation w.r.t. real time $x_0$, after a transformation from $\tau$ to real time. Similarly one may perform a change of variables in $\mathcal{W}_H$ from $\tau$ to real time $x_0$. Because the dynamical $x_0(\tau)$ can be fully absorbed into the worldline integrating factor, we no longer have any $x_0(\tau)$ dependence in the path integral, apart from the endpoints. Therefore we can cancel the $\int \mathcal{D}x_0$ factor with the implicit volume factor present in the einbein path integral measure, $\mathrm{Vol(Diff)}^{-1}$. Last due to the $0^+$ factor only paths for which $z_0>0$ contribute, which also follows from the fact that $\dot{x}_0(\tau)>0$. Gathering everything together we can find the alternate and conventional representation of the retarded propagator as
\begin{align}
    S^{\text{ret}}(x,y)&= i\theta(z_0)
   \int_y^x \!\mathcal{D}\bm{x}
   \int\frac{\mathcal{D}\bm{k}}{2\pi}\,\e^{\frac{i}{\hbar}S_{H}}\notag\\
   &\quad\times\mathcal{P}\e^{\frac{i}{\hbar}\int^{x_0}_{y_0}\!dx_0[  
    -v_F^{-1}\Delta\beta -\hbar\bm{k}\cdot\bm{\alpha}]}\gamma_0\,.
\end{align}
with $S_H$ given in~\eqref{eq:SH_qm}. While one could analyze the simpler form above, to bring about an emergent metric it proves convenient to keep the formulation with $x_0$ and $k_0$ unevaluated. Last, in the same way as above one can find for the advanced propagator an overall negative sign, time reversed arguments with $\theta(-z_0)$, and an anti-time path ordering $\bar{\mathcal{P}}$. Then one can find the standard quantum mechanical evolution operator as $S^{\text{ret}}(x,y)-S^{\text{adv}}(x,y)$.

Before applying the adiabatic theorem to the various propagators, let us mention a distinct merit of studying the advanced/retarded propagators in contrast to the Feynman propagator. This is due to the fact that diagonalizing $\slashed{p}\gamma_0$ (effectively $H$ within the path integral) yields the eigenvalues $\pm p_0 \gamma_0$, which stands in contrast to diagonalizing $\slashed{p}$ with eigenvalues $\pm p\gamma_0$. The gap for the advanced/retarded propagators is wider than it is since on-shell $p_0^2>p^2$. And the gap dictates, along with $\hbar$, the expansion parameter for the adiabatic theorem, making the advanced/retarded propagators a better defined quantity with which to study the adiabatic theorem and QGT. Nevertheless, there are advantages to studying both as reasoned above. Let us first treat the Feynman propagator showing its emergent spacetime from the QM before going onto the advanced/retarded propagators.

\section{QM and emergent spacetime in Feynman propagator}
\label{sec:Feynman}

\subsection{(3+1)-dimensions with \texorpdfstring{$\bm{\gamma_0}$}{γ0} diagonalization}
\label{sec:gamma0}

Let us begin our discussion by applying the adiabatic theorem to the case of a worldline Dirac spinor, which leads to Berry's phase at $\mathcal{O}(\hbar)$, and we will show, to the QM at $\mathcal{O}(\hbar^2)$. The effective Hamiltonian of the phase space worldline action is $\slashed{p}$, and is non-Hermitian. Its degenerate eigenvalues are $\pm p$ where $p \coloneqq \sqrt{ p^a p_a}$, and due to the constraint~\eqref{eq:p_constraint} $p$ is positive definite. The standard notation~\cite{Peskin:1995ev} for positive and negative eigenspinors are $\slashed{p}u_{i}=pu_{i}$ and $\slashed{p}v_{i}=-pv_{i}$ for $p_0>0$. However, here we must treat eigenspinors $p_0<0$ in an appropriate way ($p_0\neq 0$ for all paths since $p^2>0$), before having taken the residue in $p_0$ for the Feynman prescribed propagator. Therefore we use the mixed notation in which $\slashed{p}u_{i}=\mathrm{sgn}(p_0)pu_{i}$ and $\slashed{p}v_{i}=-\mathrm{sgn}(p_0)pv_{i}$, where for brevity we go ahead and write $\epsilon \coloneqq \mathrm{sgn}(p_0)$. And we normalize to $\bar{u}_{i}u_{j}=-\bar{v}_{i}v_{j}=\delta_{ij}$ and our resolution of the identity reads $\bar{u}_{i}v_{j}=0$, and $\sum_{i}u_{i}\bar{u}_{i}-v_{i}\bar{v}_{i}=I_4$. Berry's phase on the phase space worldline takes on a more complicated structure than the standard one-particle Dirac Hamiltonian of quantum mechanics, i.e., $\slashed{p}\gamma_0-mc\gamma_0$, however, the treatment we employ here leading to an emergent spacetime can only be had by diagonalizing the manifestly Lorentz invariant $\slashed{p}$. 

Consider Eq.~\eqref{eq:Wilson_Dirac} written as an infinite product of
\begin{equation}
    \mathcal{W}=\lim_{N\rightarrow\infty}\mathcal{P}\prod_{n=0}^{N-1}
   \Bigl[I_4+\frac{i}{\hbar}\frac{\tau}{N}e(\tfrac{n\tau}{N})p_{a}(\tfrac{n\tau}{N})\gamma^{a}\Bigr]\,, 
   \label{eq:path_break_up}
\end{equation}
whose adiabatic properties unfold after inserting complete sets of states in between each element. Intuitively this is carried out by inserting $\sum_{i}u_{i}\bar{u}_{i}-v_{i}\bar{v}_{i}=I_4$, evaluated at each instantaneous propertime $n\tau/N$. Using our Weyl representation we can write for our eigenvectors
\begin{align}
   u_{i}&=\frac{1}{2\sqrt{p(|p_0|+p)}}\!\begin{array}{c}
   \begin{pmatrix}(p_a\sigma^a+\epsilon p)\xi_i\\ 
   (p_a\bar{\sigma}^a+\epsilon p)\xi_i
   \end{pmatrix}\end{array}\,, \notag\\
   v_{i}&=\frac{1}{2\sqrt{p(|p_0|+p)}}\!\begin{array}{c}
   \begin{pmatrix}(p_a\sigma^a+\epsilon p)\eta_i\\ 
   -(p_a\bar{\sigma}^a+\epsilon p)\eta_i
   \end{pmatrix}\end{array}\,,  
   \label{eq:eigenspinors}
\end{align}
we use $\xi_{1}=\eta_{1}=[1,0]^{T}$ and $\xi_{2}=\eta_{2}=[0,1]^{T}$. The eigenspinors may be cast into a similarity transform in $\textrm{Spin}(3,1)$
as $s=(1/\sqrt{2})[u_{1},u_{2},v_{1},v_{2}](\gamma_{0}-\gamma_{5})$ that is evaluated to
\begin{equation}
   s=\frac{\slashed{p}\gamma_0+\epsilon p}{\sqrt{2p(p+|p_0|)}}\,,\quad 
   s^{-1}=\frac{\gamma_0\slashed{p}+\epsilon p}{\sqrt{2p(p+|p_0|)}}\,.
   \label{eq:sim_transform}
\end{equation}
The merit of using the mixed eigenspinor notation is now apparent in that that is no singular behaviour in the denominator of the similarity transform, as it stays positive for any $p_0$ in the path integral. We can now see that the eigenvalues of $\slashed{p}$ follow as
\begin{equation}
    s^{-1}\slashed{p}s=\epsilon p\gamma_{0}\,,
    \label{eq:gamma_0}
\end{equation}
from the similarity transform. Also we can determine the resolution of identity of the similarity transform is simply $\sum_{i}u_{i}\bar{u}_{i}-v_{i}\bar{v}_{i}=ss^{-1}$. Last, the similarity transform is not unique and any $s\to su$ for $u\in$ SO$(3)$ with $[u,\gamma_0]=0$, the little group of $\textrm{Spin}(3,1)$, will lead to the same eigenvalues. This is due to the eigenvalue degeneracy and the non-Abelian Berry phase of (3+1)-dimensional Dirac system. 

We are now in a position to determine Berry's phase and the defining QGT. Inserting $ss^{-1}$ at each piecewise propertime point in Eq.~\eqref{eq:path_break_up} leads to the appearance of an exact Berry's phase--a pure gauge transformation of the Hamiltonian $\slashed{p}$~\cite{Copinger:2022jgg}, viz. $\slashed{p} \rightarrow s^{-1}\slashed{p}s+i\hbar s^{-1}\dot{s}$. The pure gauge Berry's phase is $B_\textrm{PG}^a=-i s^{-1}\partial^a s$, which contains both the adiabatic Berry's phase and corrections to it, $B_\textrm{PG}^a=B^a+\delta B^a$. Intuitively, the adiabatic Berry's phase is the portion of the exact phase that commutes with the `diagonal' (here $\gamma_0$) element, and other `off-diagonal' elements are exponentially suppressed according to the adiabatic theorem in the interaction picture, and let us confirm this. The similarity transform can be decomposed as $(1/\sqrt{2})[u_{1},u_{2},0,0](\gamma_{0}-\gamma_{5})=\frac{1}{2}s(I+\gamma_{0})$ and $(1/\sqrt{2})[0,0,v_{1},v_{2}](\gamma_{0}-\gamma_{5})=\frac{1}{2}s(I-\gamma_{0})$. Then the adiabatic Berry's phase is the portion in which only the elements $-i\bar{u}_i\partial_au_j$ and $-i\bar{v}_i\partial_av_j$ be kept. Gathering the pieces in question together we have for the adiabatic Berry's phase
\begin{align}
    B^a &= -\frac{i}{2}(s^{-1}\partial^{a}s+\gamma_{0}s^{-1}\partial^{a}s\gamma_{0})\\
    & =-\frac{1}{2p(|p_{0}|+p)}\varepsilon^{0abc}\Sigma_{c}p_{b}\,,
    \label{eq:Berry_31F}
\end{align}
where $\Sigma^a=\mathrm{diag}[\sigma^a,\sigma^a]$ with spatial $a$ indices and with $\sigma^a$ being the Pauli matrices. One has that $\Sigma^{a}=\frac{i}{4}\varepsilon^{0abc}[\gamma_{b},\gamma_{c}]$ and $[\gamma^a,\gamma^b]+\gamma_0 [\gamma^a,\gamma^b]\gamma_0=4i\varepsilon^{0abc}\Sigma_c$. Only spatial indices are summed over in $\Sigma^a$. We also have for the non-adiabatic parts, which anti-commute with $\gamma_0$,
\begin{equation}\label{eq:delB}
    \delta B_a =i\frac{\epsilon}{2p}\Bigl\{\frac{\delta_{a}^{0}\epsilon+\hat{p}_{a}}{|p_{0}|+p}p^{i}-\delta_{a}^{i}\Bigr\}\gamma_{i}\gamma_{0}\,,
\end{equation}
and $\hat{p}_{a}\equiv p_{a}/p$. The non-Abelian curvature associated with the adiabatic Berry's phase is defined as $S^{ab}=\partial^a B^b-\partial^bB^a+i[B^a,B^b]$, and reads
\begin{equation}\label{eq:Berry_S_31F}
    S^{ab}=\frac{\hat{p}^{[a}+\eta^{[a0}\epsilon}{p^{2}(|p_{0}|+p)}\varepsilon^{0b]cd}p_{c}\Sigma_{d}+\frac{1}{2p^{2}}\varepsilon^{0abc}\Sigma_{c}\,;
\end{equation}
our antisymmetrized indices $[a,b]$ include a $1/2$ normalizing factor. Let us mention in passing that one has the radial constraint that $p_aB^a=p_a\delta B^a=p_aS^{ab}=0$. And under a gauge transformation one has $B_a\to u^\dagger B_a u -iu^\dagger \partial_a u$, $\delta B_a\to u^\dagger \delta B_a u$, and $S_{ab}\to u^\dagger S_{ab} u$. The covariant Berry phase--and we will go on to show for the QM as well--we have shown above differs from the conventional one appearing in quantum mechanics due primarily to the fact that we are applying the adiabatic theorem \textit{before} going on-shell; we will revisit this comparison when applying the adiabatic theorem to the advanced/retarded propagators.

The Berry phase in~\eqref{eq:Berry_31F} is non-Abelian, reflecting the twofold spin degeneracy of each eigenspace of $\slashed{p}$, and therefore describes parallel transport within the corresponding degenerate spin subspace. Its radial constraint, $p_{a}B^{a}=p_{a}S^{ab}=0$, shows that the connection is insensitive to changes in the overall momentum scale and instead probes changes in the direction of the timelike momentum. Consequently, transport around a closed trajectory in momentum space generally produces a path-ordered little-group rotation rather than an Abelian phase, with the associated curvature in~\eqref{eq:Berry_S_31F} determining the resulting holonomy. The connection has a monopole-like momentum-space structure, with the degeneracy at $p=0$ playing the role of its singular point; however, since here we require only $p>0$ and have not restricted the trajectory to a fixed mass shell, no quantized topological charge follows directly from~\eqref{eq:Berry_S_31F}. Rather, on each connected timelike region away from $p=0$ the bundle 
may be chosen smoothly, while its nonvanishing curvature retains the local geometric information responsible for spin-dependent transport.
\begin{figure} 
    \centering
    \includegraphics[width=\linewidth]{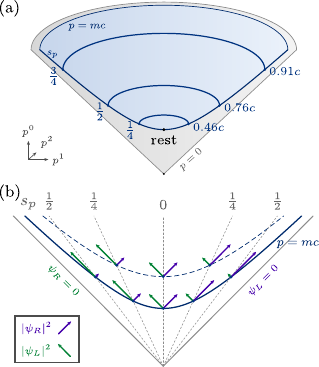} 
    \caption{Quantum metric~\eqref{eq:qgm2} on the future cone of momentum space, $p^3$ suppressed. \textbf{(a)} Mass shell $p=mc$ inside the light cone $p=0$, cut along $p^2 = 0$, with contours of constant quantum distance $s_p$ from rest.
    \textbf{(b)} In the cut plane, the dashed curve shows a sheet of constant $p\ne mc$, and purple and green arrow lengths represent the right- and left-handed probabilities $|\psi_{R,L}|^2/(\psi^\dagger\psi)=\tfrac12(1\pm p^1/p^0)$ for spin along $+p^1$.}
    \label{fig:quantum_metric}
\end{figure}

Let us now determine the QGT for our non-Abelian Berry's phase~\cite{Ma:2010ovr}. It is defined as
\begin{equation}\label{eq:Q}
    Q_{ab}=\mathfrak{g}_{ab}-\frac{i}{2}S_{ab}\,.
\end{equation}
For our non-Hermitian Hamiltonian and treatment with Weyl matrices one may show that the QGT for our system may be decomposed into spinors as
\begin{equation}
    Q_{ab}=-\frac{1}{2}\gamma_{0}(\gamma_{0}-\gamma_{5})\begin{pmatrix}\!\partial_{a}\bar{U}(1-U\bar{U})\partial_{b}U\;,\;0\\
   0\;,\;\partial_{a}\bar{V}(1+V\bar{V})\partial_{b}V
\end{pmatrix}(\gamma_{0}-\gamma_{5})\,,
\end{equation}
where $U$ and $\bar{U}$ denote the matrices $U=(u_1,u_2)$ and $\bar{U}=(\bar{u}_1,\bar{u}_2)^T$ and likewise for $V$ with $\nu$. We then have for the curvature that $S_{ab}=i(Q_{ab}-Q_{ba})$. Note that in our conventions the projection operators with $V$ acquire a minus sign. And for the QGM we have that $\mathfrak{g}_{ab}=(1/2)(Q_{ab}+Q_{ba})$. To determine the quantity it is convenient to first recast into the similarity transform; then we find that
\begin{align}
    \mathfrak{g}_{ab}&=-\frac{1}{2}\partial_{(a}s^{-1}\partial_{b)}s-\frac{1}{2}\gamma_{0}\partial_{(b}s^{-1}\partial_{a)}s\gamma_{0}+B_{(a}B_{b)}\\
    &=-\delta B_{(a}\delta B_{b)}\,.\label{eq:qgm}
\end{align}
Using~\eqref{eq:delB} let us evaluate each contribution to the quantum metric explicitly to
\begin{align}
    \mathfrak{g}_{00}&=\frac{\boldsymbol{p}^{2}}{4p^{4}}\,,\quad
    \mathfrak{g}_{ij}=-\frac{1}{4p^{2}}[\eta_{ij}-\hat{p}_{i}\hat{p}_{j}]\,,\notag\\
    \mathfrak{g}_{0i}&=\frac{1}{4p^{4}}p_{0}p_{i}\,.
\end{align}
Furthermore we can gather all together in compact Lorentz index form 
\begin{equation}\label{eq:qgm2}
    \mathfrak{g}_{ab}=-\frac{1}{4p^{2}}\bigl(\eta_{ab}-\hat{p}_{a}\hat{p}_{b}\bigr)\,.
\end{equation}
The line element $ds_p^2=\mathfrak{g}_{ab}dp^adp^b=-\tfrac14d\hat p_a d\hat p^a$ describes the hyperbolic geometry of normalized timelike momenta depicted in Fig.~\ref{fig:quantum_metric}. The geodesic distance from rest is half the rapidity, $s_p=\tfrac12\operatorname{arctanh}(v/c)$, and diverges at the light cone. The metric measures changes in the ratio of the right- and left-handed amplitudes under boosts, with one component vanishing in the light-cone limit, so the radial rescaling leaves the eigenspinor unchanged, $\mathfrak{g}_{ab}p^b=0$.

\subsubsection{\texorpdfstring{${\mathcal{O}(\hbar^2)}$}{O(h2)} expansion in path integral}

Having explored the various quantities related to Berry's phase as they pertain to our setup, let us see how the quantities emerge within our path integral. Let us begin with the exact pure gauge case. Formally expressing the path ordering exponential in its infinite product form, eq.~\eqref{eq:path_break_up}, then insert $ss^{-1}=I_4$ in between each element, and finally take the $N\to\infty$ continuum limit to find
\begin{equation}
    \mathcal{W}=s_1\mathcal{P}\e^{\frac{i}{\hbar}\int^1_0 d\tau\bigl[ \epsilon \gamma_0 pe -\hbar B_{\textrm{PG}a}\dot{p}^a \bigr]}s^{-1}_0\,.
\end{equation}
As the above stands it is exact. No adiabatic approximation has been performed. Here $s_1\equiv s(p(1))$ and likewise $s^{-1}_0\equiv s^{-1}(p(0))$.

We analyze the effects of Berry's phase not just to the leading order (here in $\hbar$) but using the systematic procedure as developed in~\cite{Mameda:2025rfn} we treat the next-to-leading order contribution at $\mathcal{O}(\hbar^2)$. This is accomplished with another transformation that serves to kill leading order contributions from the non-adiabatic part, namely it is
\begin{equation} 
    [U,\epsilon \gamma_{0}pe]=\hbar \delta B^{a}\dot{p}_{a}\,,
\end{equation}
where $U=\hbar \epsilon \delta B^{a}\dot{p}_{a}\gamma_0/(2pe)$ 
And the transformation matrix is $\exp(U)$. Performing it again on the path ordered expression we find
\begin{equation}
    \mathcal{W}=s_1\e^{-U(p(1))}\mathcal{P}\e^{\frac{i}{\hbar}\int^1_0 d\tau\,\mathcal{L}_\mathcal{W}}\e^{U(p(0))}s^{-1}_0\,,
\end{equation}
where the term in the action becomes to $\mathcal{O}(\hbar^{2})$
\begin{align}
    \mathcal{L}_\mathcal{W}&= \e^{U}\{\epsilon \gamma_0 pe -\hbar B_{\textrm{PG}}^a\dot{p}_a \}\e^{-U} +i\hbar \e^{U}\frac{d}{d\tau}\e^{-U}\\
    &=\epsilon\gamma_{0}pe-\hbar B^{a}\dot{p}_{a}-i\hbar\dot{U}\notag\\
    &\quad+\frac{\hbar^{2}\epsilon}{2pe}\Bigl\{\delta B^{(a}\delta B^{b)} - [\delta B^{a},B^{b}]\Bigr\}\gamma_{0}\dot{p}_{a}\dot{p}_{b}\,.
    \label{eq:LWhbar2}
\end{align}
First thing we can notice is that there is a new `geometric phase,' yet to $\mathcal{O}(\hbar^2)$ it represents a total, and convenient, derivative, $\dot{U}$. This is because under the path ordering after taking the total derivative it serves to cancel the enclosing $\exp(U(p(1)))$ and $\exp(-U(p(0)))$ factors. Importantly, notice in the final expression above the presence of the quantum metric,~\eqref{eq:qgm}, since $-\delta B^{(a}\delta B^{b)}=\mathfrak{g}^{ab}$. We do have one additional term $[\delta B^{a},B^{b}]$ that does not commute with $\gamma_0$. Under the adiabatic theorem, however, only parts of the action in the path ordered expression commuting with the diagonal element, here $\gamma_0$, will remain. This is because the path ordered element may be recast into an interaction picture with a dynamical phase, $\pm (i/\hbar)\int^1_0 d\tau \gamma_0 pe$, serving to suppress the off-diagonal parts through an acquired rapidly oscillating phase. The diagonal parts (those commuting with $\gamma_0$) are unaffected. Applying the adiabatic theorem we find for the path ordered expression, 
$\mathcal{W}=s_1\mathcal{P}\exp[(i/h)\int^1_0 d\tau\,\mathcal{L}_\mathcal{W}]s^{-1}_0$, where now the portion of action reads
\begin{equation}\label{eq:L_wgamma0}
    \mathcal{L}_\mathcal{W}=\epsilon \gamma_{0}pe-\hbar B^{a}\dot{p}_{a} -\frac{\hbar^{2}\epsilon}{2pe}\dot{p}^a\mathfrak{g}_{ab}\dot{p}^b\gamma_0\,.
\end{equation}

Let us next define the following projection operators: $P^\pm_{\epsilon}=(1/2)(1\pm \epsilon \gamma_0 )$, which take $\epsilon \gamma_0P^\pm_\epsilon=\pm P^\pm_\epsilon$. 
However, let us caution that the projection operators here do not correspond to positive and negative energy ones. This is because the einbein is not enforcing the mass-shell criteria; one has either $mc=\pm p$. Using the projection operators we may break up $\mathcal{W}$ into the sum of both projections. First, let us allow the projection operators to act on the similarity transform where one finds
\begin{equation}\label{eq:Pepsilonrel}
    s_1 P^\pm_\epsilon =\frac{1}{2}\bigl( 1\pm \slashed{\hat{p}}(1)  \bigr)s_1\eqqcolon P^\pm_1 s_1\,.
\end{equation}
We remark here that in the on-shell limit, with $\slashed{p},p\to mc$, $P^-$ vanishes. Then we can express the path ordered element as $\mathcal{W}=\mathcal{W}^++\mathcal{W}^-$ where
\begin{equation}\label{eq:Wpm0}
    \mathcal{W}^\pm=P^\pm_1 s_1 \mathcal{P}\e^{\frac{i}{\hbar}\int^1_0 d\tau\,\mathcal{L}_\mathcal{W}^\pm}s^{-1}_0\,,
\end{equation}
with element within the action now finally reading
\begin{equation}\label{eq:L_F_final} 
    \mathcal{L}_\mathcal{W}^\pm=\pm pe-\hbar B^a\dot{p}_{a}\mp\frac{\hbar^{2}}{2pe}\dot{p}^a\mathfrak{g}_{ab}\dot{p}^b\,.
\end{equation}
We stress that at this point apart from the constraint of $p>0$ under the path integral and an $\mathcal{O}(\hbar^2)$ truncation as is appropriate in the adiabatic approximation, no further approximations have been made.

Let us confirm at this point that the free propagator is reproduced in the field-free limit with $A\to 0$. In such a case the Berry phase and QM vanish, as does the path ordering, and the path integrals too collapse with one remaining integral in momentum space. We also select for the einbein gauge the constant $e=T$ leading to the Schwinger propertime representation. What we find is that the entire propagator reduces to
\begin{align}\label{eq:free_Ppm}
    S^c(x,y)&=i\int \!\frac{d^4p}{(2\pi\hbar)^4}\int^\infty_0\!dT\,\e^{\frac{i}{\hbar}[-(mc-i\epsilon)T-p\cdot z]}\notag\\
    &\quad \times[P^{+}\e^{ipT}+P^{-}\e^{-ipT}]\,.
\end{align}
Carrying out the $T$ integral and gathering terms we can find that indeed the propagator reduces to the free field case with
\begin{equation}
    S(x,y)=-\int\!\frac{d^{4}p}{(2\pi\hbar)^{4}}\frac{\slashed{p}+m}{p^{2}-m^{2}+i\epsilon}e^{-\frac{i}{\hbar}pz}\,.
\end{equation}
Thus we can see that in this representation the einbein no longer projects the mass-shell condition for the regular projection, $P^-$. Further at this point note that due to our demand of~\eqref{eq:p_constraint} and with the semi-classical criteria, one may actually reproduce the free propagator from~\eqref{eq:free_Ppm} with just $P^+\simeq (1/2)(1+\slashed{p}/mc)$ and $p,\slashed{p}\simeq mc$ under the integral with $P^-\simeq 0$. This fact points to an important simplification of the Feynman propagator.

Encompassing our aim to the full background gauge field dependent case we can see a similar logic applies. Namely for small $\hbar$ and with~\eqref{eq:p_constraint} we do not treat off-shell $p^2$ and hence steepest descents in $e$ demands that paths over $P^+$ are dominant to those projected with $P^-$, which instead are fixed with the positive definite and rapidly oscillating phase of $mc+p$ and are suppressed. Furthermore, the paths about $P^+$ contain the physically relevant mass-shell $p=mc$ through the Lagrange multiplier $e$. Therefore at this point let us confine our attention to the $P^+$ projected part of~\eqref{eq:L_F_final}. Also, $P^+$ when translated into its operator form using canonical momenta appears as the prefactor sitting in the conventional quadratic representation of the spinor propagator, i.e. as $i\hbar\slashed{D}+mc$. Let us next analyze precisely what the new Lagrangian means in terms of its symplectic structure and curved momentum space.

The radial identity $\mathfrak g_{ab}p^b=0$ shows that the quantum metric measures the changes in momentum direction. Writing $\hat p^a = p^a /p$, its line element reads $ds_p^2 = -\tfrac{1}{4}\eta_{ab} d\hat p^a d\hat p^b$. On each sheet of $d\hat p^a d\hat p_a = 1$, this is the positive metric of $H^3$ and the radial momentum scale remains a null direction of the 4d tensor.
For the non-Abelian $P^+$ sector, introduce a two-component spin amplitude $z$, normalized by $z^\dagger z =1$ and defined up to an overall phase~\cite{Copinger:2022jgg}. In the diagonal frame, let $B_a$ and $S_{ab}$ denote their restrictions to the retained doublet. With $Dz=dz+iB_a dp^a$, the first order part of the action has the phase-space one-form
\begin{equation}
    \Theta_0 = -(p_a + q A_a)dx^a + i \hbar z^\dagger Dz.
\end{equation}
Its exterior derivative contains the Berry-curvature contribution $-\frac{\hbar}{2}z^\dagger S_{ab}z dp^a \wedge dp^b$, together with electromagnetic and spin terms, the spin variables are retained for non-Abelian transport.

The quadratic momentum-velocity term also contributes to the reduced symplectic structure at $\mathcal{O}(\hbar^2)$~\cite{Mameda:2025rfn}. On solutions admitting a regular semiclassical expansion, up to $\mathcal{O}(\hbar)$ we have $\dot p^a \simeq v_0^a$ where $v_0^a = qe F^a{}_{b}\hat p^b$. Order reduction of this term yields
\begin{equation}
    \Theta_{\rm red} = \Theta_0 -\hbar^2 \kappa_a dp^a
\end{equation}
where $\kappa_a = \tfrac{q}{p^2}\mathfrak{g}_{ab} F^{b}{}_{c} p^c$ which, for the covariant metric, becomes $-qF_{ab}p^b / (4p^4)$. The accompanying worldline Hamiltonian is
\begin{equation}
    \mathcal{H}_{\rm red} = -e(p-mc) - \frac{\hbar^2}{2pe}\mathfrak{g}_{ab}v_0^a v_0^b .
\end{equation}
Thus $d\Theta_{\rm red}$ includes both a momentum-space correction and, through the position dependence of $\kappa_a$, mixed position-momentum dependent terms. Its pullback to a gauge-fixed constraint surface gives the physical symplectic form. The inverse defines the Dirac brackets, and the associated volume form which supplies the Liouville measure.

\subsubsection{Induced Lorentzian geometry in configuration space}
\label{sec:induced}

To obtain a leading order expression in the semiclassical limit in which $\hbar/S$ is small let us perform a semiclassical expansion for the momenta d.o.f. In the stationary phase picture to leading order we neglect contributions entering at $\mathcal{O}(\hbar)$ in the classical trajectory. Writing $\mathcal{L}_\mathcal{W}^+=\mathcal{L}_\mathcal{W}^{+ (0)}+\hbar \mathcal{L}_\mathcal{W}^{+(1)}+\hbar^2\mathcal{L}_\mathcal{W}^{+(2)}+\cdots$, and coupling to $S_A$, which enters at $\mathcal{O}(\hbar^0)$, we see that the leading order classical solution is described with $S_A+\int^1_0 d\tau \mathcal{L}_\mathcal{W}^{+ (0)}$. Here the coupled e.o.m are effectively described by the Lorentz force, 
\begin{equation}\label{eq:Lorentz}
    \dot{p}^a=qF^{ab}\dot{x}_b\,,
\end{equation}
and the velocity-momentum relation $\dot{x}_a= \hat{p}_ae + \mathcal{O}(\hbar)$ for the $P^+$ projection. We further take that the classical momentum obeys the mass-shell conditions of $m^2c^2=p^2$ and hence $mc=p$ to $\mathcal{O}(\hbar)$, consistent with the semiclassical expansion. The momentum endpoints are integrated over, while the coordinate endpoints obey $x(0)=y$ and $x(1)=x$ in accordance with the path integral definition in~\eqref{eq:Green}. The leading classical solutions provide the momenta at which we evaluate the Berry connection and the QM,
\begin{align}
    B^a\dot{p}_a&\to q B^a(p_{\textrm{cl}}(x))F_{ab}(x)\dot{x}^b\eqqcolon \mathcal{B}_a(x)\dot{x}^a\,,\label{eq:Breplace}\\
    \frac{\dot{p}^a\mathfrak{g}_{ab}\dot{p}^b}{2pe}&\to-\frac{q^2\dot{x}^aF_{ab}(x)\mathfrak{g}^{bc}(p_{\textrm{cl}}(x))F_{cd}(x)\dot{x}^d}{2mce}\notag \\
    &\eqqcolon \frac{mc}{2e}\dot{x}^a h_{ab}(x)\dot{x}^b\,,\label{eq:QGMreplace}
\end{align}
where we have suggestively written the term with the quantum metric in the form of a metric deformation.

Let us analyze the replacement of momenta d.o.f. under the path integral with~\eqref{eq:Lorentz} in a different context that better illustrates its validity. We show this through a functional integration by parts. It is sufficient to illustrate this with just the QM; let us take the following terms which are assumed to lie under the path integral measure as shown in~\eqref{eq:Green}:
\begin{equation}
    \e^{\frac{i}{\hbar}\int^1_0d\tau\,\frac{\hbar^2}{2pe}\dot{p}^a\mathfrak{g}_{ab}\dot{p}^b}\e^{\frac{i}{\hbar}\int^1_0d\tau\,[-p_a\dot{x}^a-qA_a\dot{x}^a]}\,.
\end{equation}
Then by writing the $\dot{p}^a(\tau)$ as a functional derivative of $-i\hbar\delta/\delta x^a(\tau)$ acting on the $p_a\dot{x}^a$ term as a source-like term, and then integrating by parts in the coordinate path integral one can find that the above becomes
\begin{equation}
    \e^{-\frac{i}{\hbar}\int^1_0d\tau\,p_a\dot{x}^a}\e^{\frac{i}{\hbar}\int^1_0d\tau\,\frac{-\hbar^4}{2pe}\frac{\delta}{\delta x_a}\mathfrak{g}_{ab}\frac{\delta}{\delta x_b}}\e^{-\frac{i}{\hbar}\int^1_0d\tau\,qA_a\dot{x}^a}\,.
\end{equation}
To leading order of course upon performing the functional derivative one will find that $\delta/\delta x_a\to qF^{ab}\dot{x}_b$ as employed for the $\dot{p}$ in~\eqref{eq:QGMreplace}. However, for higher orders in addition to the replacement in~\eqref{eq:QGMreplace} (to all orders in the QM) one would pick up factors of $(\delta^n/\delta x^n_{c,d,e,...})F_{ab}\dot{x}^b$. Yet, since these factors all stem from the functional derivative acting on a QM term, they will all enter with a higher power in $\hbar$ (in comparison to the derivatives acting on the $A_a\dot{x}^a$ term. Therefore in this example the replacement given in~\eqref{eq:QGMreplace} with $F_{ab}\dot{x}^b$ still representing a fluctuating variable to be integrated over is completely consistent with our $\mathcal{O}(\hbar^2)$ truncation scheme in the action.

For a concrete solution of the Lorentz-force equations, consider a background with a single nonzero gauge-field component $A_a(x^b)$, where $a\neq b$. An example is a spatially inhomogeneous magnetic field in the $\hat x_3$ direction generated by $A_1(x^2)$. Since the background is independent of $x^a$,~\eqref{eq:Lorentz} gives the conserved canonical momentum
\begin{equation}
    p^b_\textrm{cl}(x(\tau))=-qA^b(x(\tau))+p^b_0
\end{equation}
with no sum over $a$. The integration constants specifying the classical solution are implicit in $p_{\rm cl}(x)$.

Having evaluated the momenta in Berry's phase and the QM according to~\eqref{eq:Breplace} and~\eqref{eq:QGMreplace}, let us complete the remaining parts of the momentum path integral. A rigorous way to achieve this is to treat the momentum integral as an infinite product of regular integrals over time slices of width $\delta\tau$, since our action possesses no propertime derivatives in momentum. First let us introduce a positive auxiliary field $\sigma$ that gives a Fresnel representation:
\begin{equation}\label{eq:sigma_aux}
    e^{\frac{i\delta\tau}{\hbar}ep}=\sqrt{\frac{\delta\tau}{2\pi i\hbar}}\int_0^\infty\frac{e d\sigma}{\sqrt{\sigma}} \exp \bigg[\frac{i\delta\tau}{2\hbar}\Big(\frac{p^2}{\sigma}+\sigma e^2\Big)\bigg].
\end{equation}
The integration domain and square-root branch are fixed by the causal analytic continuation. With the geometric terms evaluated at their classical values, the remaining momentum integral becomes Gaussian,
\begin{equation}
    \int d^4p \exp\bigg[\frac{i\delta\tau}{\hbar}\Big(\frac{p^2}{2\sigma}-p_a\dot{x}^a\Big)\bigg]=\mathcal{C}_{\delta\tau}\sigma^2 e^{-\frac{i\delta\tau}{2\hbar}\sigma\dot{x}^2}
\end{equation}
where the field and path-independent factor $\mathcal{C}_{\delta\tau}$ is absorbed into the normalization. Together with~\eqref{eq:sigma_aux} gathering all the products, it produces an overall $\mathrm{det}(e \sigma^{3/2})$ factor and local phase,
\begin{equation}
    \Phi(e,\sigma)=-mce+\frac{\sigma}{2}(e^2-\dot{x}^2)-\frac{\hbar^2mc}{2e}\dot{x}^ah_{ab}\dot{x}^b.
\end{equation}
For timelike paths with $\dot x^2>0$, the joint $(e, \sigma)$ saddle up to $\mathcal{O}(\hbar^2)$ is, 
\begin{equation}
\begin{aligned} 
    e_\star &=\sqrt{\dot{x}^2},\qquad \sigma_\star \simeq mc[\dot{x}^a (\eta_{ab} + \hbar^2 h_{ab}) \dot{x}^b]^{-1/2},\\
    \Phi_\star &\simeq -mc\sqrt{\dot{x}^a (\eta_{ab} + \hbar^2 h_{ab}) \dot{x}^b}.
\end{aligned} \label{eq:einbein_sigma_saddle}
\end{equation}
The fluctuation determinant thus satisfies 
$|\det\Phi''_\star|^{-1/2}=\det (1/e_\star)$,
canceling the factor $e_\star$ in the prefactor. Up to a field and path-independent normalization, the regulated time-slice product yields the local measure 
$\mu[x]\coloneqq \det\sigma_\star(\tau)^{3/2}$, 
which up to $\mathcal{O}(\hbar^4)$ reads,
\begin{equation}
    \mu[x]\simeq \det
    \left \{ \left(\frac{mc}{\sqrt{\dot{x}^2}}\right)^{3/2} \left[1-\frac{3\hbar^2}{4}\frac{\dot{x}^ah_{ab}\dot{x}^b}{\dot{x}^2}\right] \right\}
\end{equation}
Restoring the electromagnetic coupling and the leading Berry transport, the causal propagator near the mass-shell pole becomes
\begin{equation}
    \label{eq:Scfinal}
    \!\!\!S^c (x,y) \simeq i
    P_x^+s_x\int_y^x \!\!\mathcal{D}x\, \mu[x] 
    \mathcal{P} e^{\frac{i}{\hbar}\int_0^1 d\tau \mathcal{L}_{\rm eff}} s_y^{-1}\,,
\end{equation}
We leave implicit the overall normalization factor, which includes the path independent Faddeev--Popov determinant.
And we find for the effective worldline Lagrangian 
\begin{equation}
    \mathcal{L}_{\rm eff} = -mc\sqrt{\dot{x}^a g^{\rm ind}_{ab}(x)\dot{x}^b}-qA_a\dot{x}^a-\hbar\mathcal{B}_a\dot{x}^a\,,
\end{equation} 
where we identify the induced effective metric as
\begin{equation}
    g_{ab}^{\rm ind}(x) \coloneqq \eta_{ab} + \hbar^2 h_{ab}(x)\,.
    \label{eq:induced_effective_metric}
\end{equation}

Integrating out the momenta in the adiabatic worldline action gives a configuration-space kinetic term with the induced metric~\eqref{eq:induced_effective_metric}, whose signature remains Lorentzian in the perturbative regime. The QM measures the distance between internal fermionic states at neighboring momenta, while the Lorentz force relates momentum changes to spacetime displacements. The resulting $F\cdot\mathfrak{g}\cdot F$ contraction pulls the momentum-space quantum metric back to configuration space along the classical trajectory.

The field-dependent part of the local measure may also be absorbed into the effective action, giving the purely imaginary contribution $\delta S_{\rm eff}=-i\hbar\,\delta\ln\mu=\mathcal{O}(\hbar^3)$, where $\delta\ln\mu$ denotes the change in $\ln\mu$ relative to the undeformed measure. This contribution modifies the propagation amplitude, while the induced metric determines the real $\mathcal{O}(\hbar^2)$ correction to the Hamilton--Jacobi phase and the pole location.

The induced metric can be treated using the geometric expansion of linearized gravity, with $\hbar^2$ playing the role of $\kappa = \sqrt{32\pi G}$. First, although trivial we can now directly interpret the metric as describing curved indices with $g_{ab}^{\rm ind}\to g_{\mu\nu}^{\rm ind}$ and $g^{\mu\nu\,{\rm ind}}\equiv \eta^{\mu\nu}- \hbar^2 h^{\mu\nu}$. Let us make this step more concrete by defining a vielbein so that $e^{\;a}_\mu=\partial x^a/\partial x^\mu$ and $g_{\mu\nu}^{\rm ind}=e^{\;a}_\mu \eta^{ab}e^{\;b}_\nu$. We also redefine our Berry phase and gauge so that $\hbar\mathcal{B}_\mu=\hbar e^{\;a}_\mu\mathcal{B}_a=\hbar \delta^{a}_\mu\mathcal{B}_a$ and $A_\mu=e^{\;a}_\mu A_a$ respectively. Upon a change of variables to the tetrad basis we are left with the effective Lagrangian
\begin{equation}\label{eq:Leff_curved}
    \mathcal{L}_{\rm eff} = -mc\sqrt{\dot{x}^\mu g^{\rm ind}_{\mu\nu}(x)\dot{x}^\nu}-qA_\mu\dot{x}^\mu-\hbar\mathcal{B}_\mu\dot{x}^\mu\,.
\end{equation} 
And the path integral measure is modified as $\int_y^x \mathcal{D}x\, \mu[x]\to\int_y^x \mathcal{D}x \, \sqrt{|g^{\rm ind}|}\mu[x]$ from the Jacobian with $|g^{\rm ind}|\equiv |\det(g_{\mu\nu}^{\rm ind})|=\prod_\tau |g_{\mu\nu}^{\rm ind}(x(\tau))|=\det (1+\hbar^2 h/2)$ with $h=\eta^{\mu\nu}h_{\mu\nu}$. The coordinate end points $x$ and $y$ too are understood in curved indices. Building on the linearized gravity identification, let us finally remark on a genuine induced geometry. The Riemann tensor is $R^{\rm ind}_{\mu\nu\alpha\beta}=(\hbar^2/2)[\partial_\alpha\partial_\nu h_{\mu\beta}+\partial_\beta\partial_\mu h_{\nu\alpha}-\partial_\beta \partial_\nu h_{\mu \alpha}- \partial_\alpha \partial_\mu h_{\nu \beta}]$ and the Ricci scalar is then $R^{\rm ind}=\hbar^2(\partial_\mu\partial_\nu h^{\mu\nu}-\eta^{\mu\nu}\partial_\mu \partial_\nu h)$, with both to $\mathcal{O}(\hbar^2)$. The Ricci scalar is generally non-vanishing for the metric perturbation,~\eqref{eq:QGMreplace}, and can contain derivatives of either the field or QM once evaluated on its classical path. We will explore some specific phenomenological examples in the context of a 3-dimensional Dirac material later. Now, however, let us show how a similar construction with differing Berry phase can be seen by not diagonalizing to $\gamma_0$, but to $\gamma_5$. We will then later address the (2+1)-dimensional Feynman case.


\subsection{(3+1)-dimensions with \texorpdfstring{$\bm{\gamma_5}$}{γ5} diagonalization}
\label{sec:gamma5}

Instead of a $s\in$ Spin(3,1) similarity transform that effectively `diagonalizes' $\slashed{p}$ on the worldline to $\epsilon \gamma_0 p$, one may equally well `diagonalize' with another $s\in$ Cl(3,1) similarity transform to $\gamma_5 p$. A distinct advantage we will go on to show is the resulting path integral maintains manifest Lorentz invariance \textit{even after adiabaticity}. We will also find the resulting expression in Berry's phase takes on simplified form in contrast to the case analyzed above. The QM remains the same in both cases.

For this section, as well as for the (2+1)-dimensional case to come, many of the steps outlined in the previous section, Sec.~\ref{sec:gamma0}, are similar. Therefore we may recycle many of the same steps, and in the process illustrate the applicability of our approach. Also, we use the same notations in each individual section to both reduce clutter and for ease in comparison from section to section. It is understood, however, that the variables introduced in each section only apply to that section.

The similarity transform to $\gamma_5 p$ we seek takes $s^{-1}\slashed{p} s=p \gamma_5$, which is easily found as
\begin{equation}\label{eq:s5}
    s^{-1}=\frac{1}{\sqrt{2}}(1+\gamma_{5}\slashed{\hat{p}})\,,\quad
    s=\frac{1}{\sqrt{2}}(1-\gamma_{5}\slashed{\hat{p}})\,.
\end{equation}
As we can see $s\in$ Cl(3,1) but $s\notin$ SO(3,1), and hence $s$ may not be decomposed into eigenspinors as employed before, c.f.~\eqref{eq:eigenspinors}. Nevertheless, a resolution of the identity, $s^{-1}s=I_4$ may be inserted into each piecewise propertime slice of the path ordered definition in~\eqref{eq:path_break_up}. Hence using~\eqref{eq:s5} for the adiabatic approximation is equally well valid a choice as was using~\eqref{eq:sim_transform}. However, while eigenvectors can be written down as a basis to~\eqref{eq:sim_transform}, since the eigenvectors do not resemble the standard eigenspinors in~\eqref{eq:eigenspinors}, we will keep all our definitions in this section in the form of the similarity transform, and not comment on their appearance in eigenvector form. 

Both Berry's phase and its off-diagonal part (the part that anti-commutes with $\gamma_5$) are readily found from~\eqref{eq:s5} as
\begin{equation}\label{eq:B5}
    B_a=\frac{1}{2p^{2}}\sigma_{ab}p^{b}\,,\quad \delta B_a = \frac{i}{2p}\bigl(\eta_{ab}-\hat{p}_{a}\hat{p}_{b}\bigr)\gamma_{5}\gamma^{b}\,.
\end{equation}
Likewise the associated curvature to Berry's phase is
\begin{equation}
    S^{ab}=\frac{1}{2p^{2}}\bigl(-\hat{p}^{a}\hat{p}_{c}\sigma^{bc}+\hat{p}^{b}\hat{p}_{c}\sigma^{ac}-\sigma^{ab}\bigr)\,.
\end{equation}
We can now see the advantage of diagonalizing to $\gamma_5$ in that we have kept a Lorentz invariant worldline Lagrangian. The cost comes with the broken subgroup Cl${}^{\textrm{odd}}(3,1)$, since we began with both even and odd number of gamma matrices in the Lagrangian, but after adiabaticity here we are left with only even combinations. As before, however, we have that the Berry phase and curvature obey a radial gauge condition with $B_ap^a=0$. Yet here the non-Abelian Berry phase may equally well be written with a Berry gauge transformation such that $B_a\to u^{-1}B_au-iu^{-1}\partial_a u$ with $u\in$ Cl${}^{\textrm{even}}(3,1)$ with $u$ now describing a momentum `boost' or `rotation;' $\delta B_a\to u^{-1}\delta B_a u$ as before.  Also, most notably with the covariant description of the Berry phase here we have a momentum `boost' connection in addition to the spatial rotation/little-group part that was only present in~\eqref{eq:Berry_31F}.

One also has a QGT associated to the Berry phase here. This is the same as~\eqref{eq:Q}, $Q_{ab}=\mathfrak{g}_{ab}-(i/2)S_{ab}$, from which we may define the curvature, $S_{ab}=i(Q_{ab}-Q_{ba})$, and QM, $\mathfrak{g}_{ab}=(1/2)(Q_{ab}+Q_{ba})$. However, here let us show the form for the QM in similarity transform form only: $\mathfrak{g}_{ab}=-\frac{1}{2}\partial_{(a}s^{-1}\partial_{b)}s-\frac{1}{2}\gamma_{5}\partial_{(b}s^{-1}\partial_{a)}s\gamma_{5}+B_{(a}B_{b)}$. We can easily evaluate the QM using~\eqref{eq:B5} to find exactly the same form as the (3+1)-dimensional $\gamma_0$ diagonalized one in~\eqref{eq:qgm2}.

Transforming the path ordered exponential in the propagator takes place similar to the $\gamma_0$ adiabatically case above. Let us therefore refer the reader there for details. Let us begin with the pure gauge form with
\begin{equation}
    \mathcal{W}=s_1\mathcal{P}\e^{\frac{i}{\hbar}\int^1_0 d\tau\bigl[  \gamma_5 pe -\hbar B_{\textrm{PG}a}\dot{p}^a \bigr]}s^{-1}_0\,,
\end{equation}
where now $s$ is given by~\eqref{eq:s5}, and $B_\textrm{PG}=B+\delta B$ using~\eqref{eq:B5}. Then we apply the adiabatic theorem to $\mathcal{O}(\hbar^2)$ by using another transformation matrix with $\exp(U)$ such that $[U,\gamma_{5}pe]=\hbar\delta B_{a}\dot{p}^{a}$ where $U=\hbar\delta B^{a}\dot{p}_{a}\gamma_{5}/(2pe)$. Retaining the terms that commute with $\gamma_5$ gives the Lagrangian~\eqref{eq:L_wgamma0} with $\epsilon\gamma_0$ replaced by $\gamma_5$ and the Berry connection given by~\eqref{eq:B5}. Then, we may apply projection operators; here we use $P_5^\pm=(1/2)(1\pm\gamma_5)$, which take $\gamma_5 P_5^\pm=\pm P_5^\pm$. Further we notice here that $s_1P_5^\pm=P_1^\pm s_1$ as it did before with $P_\epsilon^\pm$ in~\eqref{eq:Pepsilonrel}. Finally, we may express the final the path ordered exponential as $\mathcal{W}=\mathcal{W}^++\mathcal{W}^-$ where now $\mathcal{W}^\pm$ take the same form as~\eqref{eq:Wpm0} and Lagrangians retain the form~\eqref{eq:L_F_final}, with the similarity transform~\eqref{eq:s5} and connection~\eqref{eq:B5}.

At this point notice our setup is highly similar to that of the previous case with $\gamma_0$, and moreover all of the steps we used previously apply here as well. Therefore, we refer the reader to Sec.~\ref{sec:induced} for details. Notably, we keep the physically relevant and dominant $P^+$ part, and integrate out the momentum to discover the same induced geometry here as well. The key difference with the case here is that we found a Lorentz invariant Berry phase. With both the $\gamma_0$ and $\gamma_5$ (3+1)-dimensional cases evaluated, let us next turn our attention to (2+1)-dimensions.

\subsection{(2+1)-dimensions}

Let us now address the (2+1)-dimensional case of the Feynman propagator. Unlike the (3+1)-dimensional case we have no $\gamma_5$, so we are limited to a similarity transform diagonalizing to $\gamma_0$. Many of the computations to follow closely parallel the (3+1)-dimensional $\gamma_0$ case as well, so we will endeavor to be brief.

Let us begin by constructing the similarity transform, which is constructed from the eigenspinor; here similarly 
$\slashed{p}u=\epsilon pu$ and $\slashed{p}v=-\epsilon pv$, where one has orthonormality, $\bar{u}u=-\bar{v}v=1$ and $\bar{u}v=0$, and completeness, $u\bar{u}-v\bar{v}=I_2$. One may find the eigenspinors take the following form:
\begin{equation}
   u=\frac{\slashed{p}\gamma_0+\epsilon p }{\sqrt{2p(|p_0|+p)}}\xi_1\,, \quad v=\frac{\slashed{p}\gamma_0+\epsilon p }{\sqrt{2p(|p_0|+p)}}\eta_2\,.
\end{equation}
Definitions for $p$ and $\epsilon$ are the similar to the (3+1)-dimensional case with spatial dimension span reduced by one. The eigenspinors can be cast into a similarity transform such that $s=[u,v]$, which takes the form
\begin{equation}
   \label{eq:sim_transform_2}
   s=\frac{\slashed{p}\gamma_{0}+\epsilon p}{\sqrt{2p(p+|p_{0}|)}}\,,\quad s^{-1}=\frac{\gamma_{0}\slashed{p}+\epsilon p}{\sqrt{2p(p+|p_{0}|)}}\,,
\end{equation}
along with its inverse. Naturally, the similarity transform diagonalizes $\slashed{p}$ so that $s^{-1}\slashed{p}s=\epsilon p\gamma_{0}$.

With the similarity transform we can readily construct the pure gauge Berry phase, whose components are
\begin{align}
    B^{a}&=b^a\gamma^{0}\,,\quad b^a\coloneqq \frac{\varepsilon^{0ab}p_{b}}{2p(|p_{0}|+p)}\\
    \delta B_a &=i\frac{\epsilon}{2p}\Bigl\{\frac{\delta_{a}^{0}\epsilon+\hat{p}_{a}}{|p_{0}|+p}p^{i}-\delta_{a}^{i}\Bigr\}\gamma_{i}\gamma_{0}\,.
\end{align}
The QGT, similarly $Q_{ab}=\mathfrak{g}_{ab}-(i/2)S_{ab}$, is then composed of the curvature associated with the Berry phase,
\begin{equation}
    S^{ab}=-\epsilon\frac{\varepsilon^{abc}p_{c}}{2p^{3}}\gamma^{0}\,,
\end{equation}
and the QM taking the same from as the (3+1)-dimensional case of~\eqref{eq:qgm2}, but with reduced dimension.

With the close similarity to the (3+1)-dimensional case we can immediately infer the final form after performing the adiabatic approximation to the path ordered exponential to $\mathcal{O}(\hbar^2)$. We find indeed the final expression can be written as a sum over projections, $P^\pm_\tau=(1/2)[1\pm \slashed{\hat{p}}(\tau)]$, such that $\mathcal{W}=\mathcal{W}^++\mathcal{W}^-$ where $\mathcal{W}$ is given by~\eqref{eq:Wpm0}, and the first quantized Lagrangian reads
\begin{equation}\label{eq:L_F_final_21}
    \mathcal{L}_\mathcal{W}^\pm=\pm pe\mp \epsilon\hbar b^a\dot{p}_{a}\mp\frac{\hbar^{2}}{2pe}\dot{p}^a\mathfrak{g}_{ab}\dot{p}^b
\end{equation}
in analogy to the 3+1 case but in reduced dimension. Note that we do not have a path ordering here since the Berry phase is Abelian. In contrast to the (3+1)-dimensional case we find after applying the projection operators that the Berry phase acquires a positive negative energy sign factor, since $B_a\propto \gamma_0$. Both positive and negative energies are in fact contained in both projections, $P^\pm$. But as before we find only the $P^+$ projection both contains the physically relevant mass pole and is the dominant contribution of the path integral in the semi-classical limit. Therefore, let us just treat the $P^+$ projection.

The steps to an emergent Lorentzian geometry proceed very close to the (3+1)-dimensional case so let us just describe the primary differences. As before we have the replacements as given in~\eqref{eq:Breplace} and~\eqref{eq:QGMreplace} along with the mass-shell condition for the Berry phase and QM. The configuration-space reduction follows Sec.~\ref{sec:induced}, with the momentum Gaussian now contributing $\sigma^{3/2}$ instead of $\sigma^2$. Combining this factor with the auxiliary-field prefactor and fluctuation determinant gives the local measure $\mu[x]=\det\sigma_\star$, whose expansion reads
\begin{equation}
    \mu[x]\simeq \det
    \left\{ \left(\frac{mc}{\sqrt{\dot{x}^2}}\right)\left[1-\frac{\hbar^2}{2}\frac{\dot{x}^ah_{ab}\dot{x}^b}{\dot{x}^2}\right]\right\}\,.
\end{equation}
The resulting the path integral in coordinate space is precisely~\eqref{eq:Scfinal}, however with the projection operator, $P^+$ as given above, and similarity transforms given by~\eqref{eq:sim_transform_2}. And we also have for the effective Lagrangian,
\begin{equation}
    \mathcal{L}_{\rm eff} = -mc\sqrt{\dot{x}^a g^{\rm ind}_{ab}(x)\dot{x}^b}-qA_a\dot{x}^a-\hbar q\epsilon b^aF_{ab}\dot{x}^b
\end{equation} 
in (2+1)-dimensions and with the induced metric perturbation similarly written as
\begin{equation}\label{eq:h_f21}
    h_{ab}(x)= -\frac{q^2F_{ad}\mathfrak{g}^{dc}F_{cb}}{m^2c^2}\,.
\end{equation} 
Thus we see a similar induced geometry in (2+1)-dimensions as well. Finally, the avenue to curved indices proceeds on the same lines as for the 3+1 case. Having fully explored all the cases of the Feynman propagator we now address the advanced/retarded propagators.

\section{QGT and emergent spacetime in advanced/retarded propagators}
\label{sec:retarded}

Having seen how an emergent spacetime naturally follows from the adiabatic theorem due to constituents of the QGT in the Dirac propagator with Feynman pole prescription let us now show how it does so for the advanced/retarded propagators. As prefaced before diagonalization of the Dirac Hamiltonian, eq.~\eqref{eq:H31}, as it appears in standard quantum mechanics is known. And so too is the QM~\cite{Matsuura:2010vi,Provost:1980nc}. Thus much of the analysis that follows below for the advanced/retarded propagators, while new in the context of the phase space worldline formalism, has enjoyed study in a different but related context. As mentioned before, below we will confine our attention to the retarded propagator case, noting that the advanced one can be had with a simple redefinition of the propertime time/einbein contour; see~\eqref{eq:Sadv}. Also, we will work in units familiar to condensed matter in this section, in contrast to the previous one. But like above, a similar replacement as indicated in~\eqref{eq:unit_map}, is all that is needed to go to QED or QED${}_3$. Let us last further caution that $\hbar$ placements with the wave vector $\bm{k}$ are not be included in the $\hbar/S$ expansion since $\bm{p}=\hbar \bm{k}$. Let us first address the (3+1)-dimensional case.

\subsection{(3+1)-dimensions}

As before in the previous section let us begin with a discussion of the eigendecomposition and then layout of the Berry phase and QGT. The eigenvalues of the Hamiltonian of the retarded propagator, $H_{\bm k}=\hbar v_{F}\bm{k}\cdot\bm{\alpha}+\Delta\beta$ are $\pm E_{\bm{k}}$ with $E_{\bm{k}}=\sqrt{\Delta^{2}+\hbar^{2}v_{F}^{2}\bm{k}^{2}}$ being manifestly positive under the path integral. Thus for the retarded propagator we needn't use an arbitrary sign function in the energy to cover both cases. And we can thus define our positive and negative energy eigenvectors as $H_{\bm k} w_i^\pm=\pm E_{\bm{k}}w_i^\pm$, which for orthonormality such that $w^{i\dagger}_\pm w^j_\pm=\delta_{ij}\delta_{\pm\pm}$ one can write in the Weyl or chiral basis that
\begin{align}
    w_i^+&=\frac{1}{2\sqrt{E_{\bm{k}}\big[E_{\bm{k}}+\Delta\big]}}\begin{pmatrix}(E_{\bm{k}}+\Delta-\hbar v_{F}\boldsymbol{\sigma}\cdot\bm{k})\xi_{i}\\
(E_{\bm{k}}+\Delta+\hbar v_{F}\boldsymbol{\sigma}\cdot\bm{k})\xi_{i}
\end{pmatrix}\,,\\
    w_i^-&=\frac{1}{2\sqrt{E_{\bm{k}}\big[E_{\bm{k}}+\Delta\big]}}\begin{pmatrix}(E_{\bm{k}}+\Delta+\hbar v_{F}\boldsymbol{\sigma}\cdot\bm{k})\eta_{i}\\
(-(E_{\bm{k}}+\Delta)+\hbar v_{F}\boldsymbol{\sigma}\cdot\bm{k})\eta_{i}
\end{pmatrix}\,.
\end{align}
Like before one may construct a similarity (here a unitary) transform in the Weyl basis such that $u_{\bm{k}}=(1/\sqrt{2})[w_{1}^{+},w_{2}^{+},w_{1}^{-},w_{2}^{-}](\gamma_{0}-\gamma_{5})$, which takes the following compact form
\begin{equation}\label{eq:uk}
    u_{\bm{k}}=\frac{E_{\bm{k}}+\Delta+\hbar v_{F}(\boldsymbol{\alpha}\cdot\bm{k})\beta}{\sqrt{2E_{\bm{k}}\big[E_{\bm{k}}+\Delta\big]}}\,,
\end{equation}
with inverse 
\begin{equation}\label{eq:ukd}
    u_{\bm{k}}^{\dagger}=\frac{E_{\bm{k}}+\Delta-\hbar v_{F}(\bm{\alpha}\cdot\bm{k})\beta}{\sqrt{2E_{\bm{k}}\big[E_{\bm{k}}+\Delta\big]}}\,.
\end{equation}
The similarity transform that serves as the basis building block of the Berry phase takes
\begin{equation}
    u_{\bm{k}}^{\dagger}H_{\bm k}u_{\bm{k}}=E_{\bm{k}}\beta\,.
\end{equation}
Again we remark here that the similarity transform is not unique and any $u_{\bm{k}}\to u_{\bm{k}}u'$ for $u'\in$ SO(3) will lead to the same diagonal form. This is what leads to a non-Abelian Berry phase in the (3+1)-dimensional case.

To see the Berry phase and the quantum metric let us apply the adiabatic theorem to the path ordered exponential in~\eqref{eq:W_H}. We can accomplish this the same way as was done for the Feynman propagator case, namely we break up the path ordered exponential into piecewise propertime products (similar to~\eqref{eq:path_break_up}) and insert in between each element $u_{\bm{k}}^{\dagger}u_{\bm{k}}=I_4$ at the given propertime. This produces the path ordered element with $H_{\bm k}\to E_{\bm k}\beta+i\hbar u_{\bm{k}}^{\dagger}\dot{u}_{\bm{k}}$, where the exact pure gauge Berry phase separates into $-iu_{\bm{k}}^{\dagger}\nabla_{\bm k}u_{\bm{k}}=\bm{B}+\bm{\delta B}$ where
\begin{equation}
    \bm{B}=-\frac{(\hbar v_{F})^{2}}{2E_{\bm{k}}\big[E_{\bm{k}}+\Delta\big]}\bm{k}\times\bm{\Sigma}
\end{equation}
is the standard Berry phase after applying the adiabatic approximation, and 
\begin{equation}\label{eq:dB_ra}
    \bm{\delta B}=-\frac{i\hbar v_{F}}{2E_{\bm{k}}\big[E_{\bm{k}}+\Delta\big]}\Bigl\{(E_{\bm{k}}+\Delta)\boldsymbol{\alpha}\beta-\frac{\hbar^{2}v_{F}^{2}\bm{k}}{E_{\bm{k}}}(\bm{\alpha}\cdot\bm{k})\beta\Bigr\}
\end{equation}
is the correction to the Berry phase, whose square will ultimately provide us with the quantum metric. To gain a better insight into the Berry phase let us also express it in its eigenspinor form as
\begin{equation}
    \bm{B}=-\frac{i}{2}(\gamma_{0}-\gamma_{5})\begin{pmatrix}\!W^{+\dagger}\nabla_{\bm{k}} W^+\;,\;0\\
   0\;,\;W^{-\dagger}\nabla_{\bm{k}} W^-
\end{pmatrix}(\gamma_{0}-\gamma_{5})\,,
\end{equation}
for $W^{\pm}=(w^\pm_1,w^\pm_2)$ and $W^{\pm \dagger}=(w^{\pm \dagger}_1,w^{\pm \dagger}_2)^T$. Here the $(1/\sqrt{2})(\gamma_0-\gamma_5)$ factor is due to our Weyl representation choice for the Dirac matrices. And one may also find for the corresponding Berry curvature, $S_{ij}=\bm{\nabla}_{i}B_{j}-\bm{\nabla}_{j}B_{i}+i[B_{i},B_{j}]$ with $S_{k}=\frac{1}{2}\varepsilon_{kij}S^{ij}$, or simply ${\bm S}=\nabla_{\bm k}\times {\bm B}+i{\bm B}\times {\bm B}$, that
\begin{equation}\label{eq:ret_curvature}
    \bm{S}=\frac{(\hbar v_{F})^{2}}{2E_{\bm{k}}^{3}}\Bigl\{\frac{(\hbar v_F)^2\boldsymbol{\Sigma}\cdot\bm{k}}{E_{\bm{k}}+\Delta}\bm{k}+\Delta\bm{\Sigma}\Bigr\}\,.
\end{equation}
The topological structure of the Berry curvature describes a Dirac-monopole in momentum space, which is regularized by the mass gap. Indeed in the massless limit one would find a monopole of strength $1/2$ weighted by the helicity ${\bm k}\cdot {\bm \Sigma}/|{\bm k}|$. 

Let us next examine the QGT, this is defined here as well as $Q_{ij}=\mathfrak{g}_{ij}-(i/2)S_{ij}$, where we have that
\begin{align}
    &Q_{ij}=-\frac{1}{2}(\gamma_{0}-\gamma_{5})\\&\times\begin{pmatrix}\!\nabla_iW^{+\dagger}(1-W^+W^{+\dagger})\nabla_{j} W^+\;,\;0\\
   0\;,\;\nabla_iW^{-\dagger}(1-W^-W^{-\dagger})\nabla_{j} W^-
\end{pmatrix}(\gamma_{0}-\gamma_{5})\,,\notag
\end{align}
with no temporal components in contrast to the Feynman propagator case. Similar identities follow, however, including the separation into the QM (Fubini-Study metric), $\mathfrak{g}_{ij}=(1/2)(Q_{ij}+Q_{ji})$, and the Berry curvature,~\eqref{eq:ret_curvature}, given as $S_{ij}=i(Q_{ij}-Q_{ji})$. After some steps we can determine that the quantum metric is
\begin{align}
    \mathfrak{g}_{ij}&=-\delta B_{(i}\delta B_{j)}\\
    &=-\frac{(\hbar v_{F})^{2}}{4E_{\bm{k}}^{2}}\Bigl\{\delta_{ij}-\frac{\hbar^{2}v_{F}^{2}}{E_{\bm{k}}^{2}}k_{i}k_{j}\Bigr\}\,.
    \label{eq:qgm3}
\end{align}
In units of QED, this would become $\mathfrak{g}_{ij}=-(1/4)p_0^{-2}[\delta_{ij}-p_0^{-2}p_ip_j]$ with $p_0$ on-shell~\cite{Matsuura:2010vi,Provost:1980nc,Oancea:2025lfd},with which we can contrast to the QM found for the Feynman propagator in~\eqref{eq:qgm2}. Notice that the two share a similar form in the small momenta regime on-shell in which $|\mathfrak{g}_{ij}|\to|\eta_{ij}/(4m^{2}c^{2})|$. We remark that in contrast to the conventional definition of the retarded/advanced QM, and indeed our Feynman QM, we have an overall global minus sign difference, and this is due to our definition of the QGT; both QMs agree in physical content.

\subsubsection{Covariant and Hamiltonian quantum metrics}

The metrics~\eqref{eq:qgm2} and \eqref{eq:qgm3} use different normalizations and complementary subspaces. To compare them, set $M=\Delta/v_F>0$, $\boldsymbol\pi=\hbar\boldsymbol k$, and $p^0=E_{\boldsymbol k}/v_F=\sqrt{M^2+\boldsymbol\pi^2}$. Superscripts $F$ and $H$ identify the covariant and Hamiltonian constructions. Using $dp^0=\boldsymbol\pi\cdot d\boldsymbol\pi/p^0$, the full mass-shell pullback and the momentum-coordinate Hamiltonian tensor are
\begin{align}
    \mathfrak g^{F,\Sigma}_{ij}
    &=\frac{1}{4M^2}
    \left(\delta_{ij}-\frac{\pi_i\pi_j}{(p^0)^2}\right),\nonumber \\
    \mathfrak g^{H,\pi}_{ij}
    &=-\frac{M^2}{(p^0)^2}\mathfrak g^{F,\Sigma}_{ij},
    \label{eq:metric_comparison}
\end{align}
where $\mathfrak g^{H,\pi}=\hbar^{-2}\mathfrak g^{H,k}$, and $\pi_i$ are Euclidean Cartesian components. For the planar Dirac Hamiltonian, Appendix~\ref{app:bloch-metric} derives the Bloch metric and its relation to the covariant mass-shell pullback. The pullback includes the temporal and mixed components of the covariant tensor. The covariant normalization uses the Dirac bilinear $\bar\psi \chi$ whereas the Hamiltonian Berry connection uses the positive Hilbert inner product $\psi^\dagger \chi$~\cite{PhysRevD.91.025004}. With our signed convention, the positive Fubini--Study tensor is $g^{\mathrm{FS},\pi}=-\mathfrak g^{H,\pi}$.

The physical equivalence follows before projection, where the frames~\eqref{eq:sim_transform} and~\eqref{eq:uk} obey $u_{\boldsymbol k}=sT$, with
\begin{equation}
    T=\sqrt{\frac{M}{p^0}}\Lambda_+ +\sqrt{\frac{p^0}{M}}\Lambda_- -\frac{(\boldsymbol\alpha\cdot\boldsymbol\pi) \Lambda_-}{\sqrt{Mp^0}}.
\end{equation}
Therefore $u_{\boldsymbol k}\Lambda_+ =\sqrt{M/p^0}\,s\Lambda_+$, so the positive-energy subspaces coincide, while the last term converts their complements. For constant $M$ and dots denoting physical-time derivatives,
\begin{equation}
    T^{-1}\dot T
    =-\frac{\dot p^0}{2p^0}\beta -\frac{1}{M}\left(\boldsymbol\alpha \cdot\dot{\boldsymbol\pi}-\frac{\dot p^0}{p^0}\boldsymbol\alpha\cdot\boldsymbol\pi \right)\Lambda_-.
\end{equation}
Both the normalization derivative and the coupling between complementary subspaces would be lost by treating $T$ as constant.

For a prescribed momentum history, express the physical-time Dirac equation in the two frames, $\psi=sa=u_{\boldsymbol k}b$. Since $a=Tb$, the chain rule gives exactly
\begin{equation}
\begin{aligned}
    K_s&=s^{-1}H_{\boldsymbol k}s-i\hbar s^{-1}\dot s,\\
    K_H&=T^{-1}K_sT-i\hbar T^{-1}\dot T\\
    &=E_{\boldsymbol k}\beta
    -i\hbar u_{\boldsymbol k}^{\dagger}\dot u_{\boldsymbol k}.
\end{aligned}
\end{equation}
Their evolution operators satisfy $\mathcal U_H=T_f^{-1}\mathcal U_sT_i$, giving the same amplitudes with corresponding endpoint states and the same eigenphases for a closed history.

\subsubsection{\texorpdfstring{$\mathcal{O}(\hbar^2)$}{O(h2)} expansion in path integral}

Let us proceed as we did before by first constructing the pure gauge transformation of the path ordered exponential. This is found by breaking up the path ordering as shown in~\eqref{eq:path_break_up} and inserting $u_{\bm{k}}u_{\bm{k}}^\dagger=I_4$ in between each element. One finds~\eqref{eq:W_H} becomes
\begin{equation}
    \mathcal{W}_H=u_{\bm{k} 1}\mathcal{P}\e^{\frac{i}{\hbar}\int^1_0d\tau\,\bigl[ -v_F^{-1}E_{\bm k}\beta e -\hbar \bm{B}_\textrm{PG}\cdot  \dot{\bm{k}} \bigr]}u_{\bm{k} 0}^{\dagger}\,.
\end{equation}

Proceeding as before to apply the adiabatic theorem to $\mathcal{O}(\hbar^2)$ to the path ordered exponential we introduce another similar similarity transform, $\exp(U)$, such that $[U,-v_F^{-1}E_{\bm{k}}\beta e]=\hbar\bm{\delta B}\cdot \dot{\bm{k}}$. This is found as $U=-\hbar v_F\bm{\delta B}\cdot \dot{\bm{k}}\beta/(2E_{\bm{k}}e)$. Evaluating this transformation through $\mathcal{O}(\hbar^2)$ as in~\eqref{eq:LWhbar2} and keeping the terms that commute with $\beta$ gives two energy blocks, $\beta=\pm1$. The resulting Lagrangian has a similar gamma matrix form to the case of the Feynman propagator, so we may immediately decompose the above into positive and negative energy projections with
\begin{equation}
    P^\pm_{\bm k}=\frac{1}{2}\Bigl( 1\pm \frac{H_{\bm k}}{E_{\bm k}} \Bigr)\,,
\end{equation}
which carry some propertime dependence that we denote as before with $P_{{\bm k}1}^\pm\equiv P_{{\bm k}}^\pm({\bm{k}(1))}$. Then using the projectors we can find that the path ordered exponential decomposes into $\mathcal{W}_H=\mathcal{W}_H^++\mathcal{W}_H^-$ with 
\begin{equation}\label{eq:W_H_split}
    \mathcal{W}_H^\pm=P_{{\bm k}1}^\pm u_{{\bm k}1}\mathcal{P}\e^{\frac{i}{\hbar}\int^1_0d\tau\,\mathcal{L}_\mathcal{W}^\pm}u_{{\bm k}0}^{\dagger}
\end{equation}
and whose portion of the Lagrangian reads
\begin{equation}\label{eq:L_W_split}
    \mathcal{L}^\pm_\mathcal{W}=\mp v_{F}^{-1}E_{\bm{k}}e-\hbar\bm{B}\cdot\dot{\bm{k}}\pm\frac{\hbar^{2}v_{F}}{2E_{\bm{k}}e}\dot{k}^i\mathfrak{g}_{ij}\dot{k}^j\,.
\end{equation}
We emphasize that up to this point in the adiabatic limit the only approximation is the truncation to $\mathcal{O}(\hbar^2)$ in the path ordered exponential, with no other assumptions being made. Here, however, in contrast to the Feynman propagator case we must keep both projections, each pointing to either positive or negative energy. Neither of which is small in the semi-classical limit.

\subsubsection{Induced Lorentzian geometry in configuration space}

The fermionic retarded and advanced propagators in the field theories of condensed matter and QED with the QGT indeed have an induced Lorentzian geometry in configuration space, as was demonstrated for Feynman propagators. Again, let us assume the momenta d.o.f. in the Berry phase and QM may be evaluated on their classical expressions according to the Lorentz force e.o.m. $\hbar\dot{k}^a=qF^{ab}\dot{x}_b$, velocity momentum relation $\dot{\bm{x}}=\pm e\hbar v_{F}\bm{k}/E_{\bm{k}}+\mathcal{O}(\hbar)$, and mass-shell condition, $\hbar^2 k^2v_F^2=\Delta^2$. This will produce for the term with the Berry phase
\begin{equation}\label{eq:B_replace_ra}
    \bm{B}\cdot\dot{\bm{k}}\to -q\hbar^{-1} B^i(k_{cl}(x))F_{ia}(x)\dot{x}^a\eqqcolon \mathcal{B}_a(x)\dot{x}^a\,,
\end{equation}
and for the term with the QM
\begin{equation}\label{eq:qm_replace_ra}
    \frac{\dot{k}^i\mathfrak{g}_{ij}\dot{k}^j}{eE_{\bm k}}\rightarrow-\frac{q^2\dot{x}_aF^{ai}\mathfrak{g}_{ij}(k_{cl}(x))F^{jb}\dot{x}^b}{\hbar^2eE_{\bm k}(k_{cl}(x))}\,.
\end{equation}
The perturbation to the metric takes a different form in contrast to the Feynman propagator case we will show later. But, in the same way as for the Feynman propagator the replacement of $\hbar \dot{\bm k}$ with the Lorentz force expression is valid within our $\mathcal{O}(\hbar^2)$ expansion scheme.

Let us now complete the remaining momenta path integrals. As was demonstrated with the retarded propagator in its quantum mechanical 3-dimensional form,~\eqref{eq:Sret_def}, we can immediately take the $k_0$ integral to fix the einbein gauge with a $\delta[e-\dot{x}_0]$ factor. However since the einbein, $e$, is only valid for $e(\tau)>0$ we are restricted to positive $\dot{x}_0$ and endpoint arguments such that $x_0>y_0$, implemented with a global $\theta(z_0)$ factor. We can then address the remaining momenta integrals,
\begin{equation}
    \label{eq:3k_integrals}
    \int\frac{\mathcal{D}\bm{k}}{2\pi}\e^{\frac{i}{\hbar}\int_{0}^{1}d\tau[\hbar\bm{k}\cdot\dot{\bm{x}}\mp v_{F}^{-1}\dot{x}_0E_{\bm k}]}\,.
\end{equation}
In the small $\hbar$ limit we can evaluate this with steepest descents. Classical paths follow $\hbar\bm{k}_{cl}=\pm\dot{\bm{x}}v_{F}^{-1}\Delta(\dot{x}_0^{2}-\dot{\bm{x}}^{2})^{-1/2}$. This will send the classical action to $\mp v_{F}^{-1}\Delta\sqrt{\dot{x}_0^{2}-\dot{\bm{x}}^{2}}$. Fluctuations about the classical path will produce the functional determinant 
\begin{align}
    \mathcal{M}[x]&\coloneqq\det{}^{-1/2}\Biggl[\frac{\sqrt{\dot{x}_0^{2}-\dot{\bm{x}}^{2}}}{v_{F}^{-1}\Delta}\Bigl\{\delta^{ij}-\frac{1}{\dot{x}_0^{2}}\dot{x}^{i}\dot{x}^{j}\Bigr\}\Biggr]\notag\\
    &=\det{}\Biggl[\frac{\Delta^{3/2}}{v_F^{3/2}}\frac{|\dot{x}_0 |}{(\dot{x}_0^{2}-\dot{\bm{x}}^{2})^{5/4}}\Biggr]\,. 
\end{align}
We remark that the fluctuation determinant is independent of either the Berry phase or QM, but is present in the free field case. The same momentum Hessian appears in the scalar propagator, whose pole residue contributes an additional weight proportional to $E_k^{-1}$. Gathering everything together we find~\eqref{eq:3k_integrals} becomes
\begin{equation}
    \mathcal{M}[x]\e^{\frac{i}{\hbar}\int^1_0d\tau\,[\mp v_{F}^{-1}\Delta\sqrt{\dot{x}_0^{2}-\dot{\bm{x}}^{2}}]}\,.
\end{equation}
Last, on the classical path one can find that 
\begin{equation}
    \frac{1}{|\dot{x}_0|}=\frac{\Delta}{E_{\bm{k}}\sqrt{\dot{x}_0^{2}-\dot{\bm{x}}^{2}}}\,.
\end{equation}
Then evaluating $E_{\bm k}$ on the classical path as done in~\eqref{eq:qm_replace_ra} we again suggestively write that
\begin{equation}
    -\frac{q^2\dot{x}_aF^{ai}\mathfrak{g}_{ij}F^{jb}\dot{x}^b}{2\hbar^2v_F^{-1}\dot{x}_0E_{\bm k}}\eqqcolon
    -\frac{v_F^{-1}\Delta}{2\sqrt{\dot{x}_0^2-\dot{\bm{x}}^2}}\dot{x}^ah_{ab}(x)\dot{x}_b
\end{equation}
where now we express the perturbation to the metric of the retarded propagator as
\begin{equation}\label{eq:h_ar}
    h_{ab}= \frac{q^2v_F^2}{\hbar^2E_{\bm k}^2}F_{ai}\mathfrak{g}^{ij}F_{jb}\,.
\end{equation}

Let us go ahead and express the entire propagator now evaluated in configuration space as
\begin{align}
    S^\textrm{ret}(x,y)&=i\theta(z_0)\sum_{r=\pm} P_{\bm{k}1}^r u_{\bm{k}1}\nonumber \\
    &
    \quad
    \times\int^x_y\!\mathcal{D}x\mathcal{M}[x]\,
    \mathcal{P}\e^{\frac{i}{\hbar}\int^1_0d\tau\,\mathcal{L}_{\rm eff}^{(r)}}u_{\bm{k}0}^\dagger\gamma_0\,,
    \label{eq:S_ret_final}
\end{align}
with the effective worldline action becoming to $\mathcal{O}(\hbar^2)$
\begin{equation}\label{eq:Lr_eff}
    \mathcal{L}_{\rm eff}^{(r)} = -rv_F^{-1}\Delta\sqrt{\dot{x}^a g^{\rm ind}_{ab}(x)\dot{x}^b}-qA_a\dot{x}^a-\hbar\mathcal{B}_a\dot{x}^a\,,
\end{equation}
where the induced metric retains the definition~\eqref{eq:induced_effective_metric}, with $h_{ab}$ now given by~\eqref{eq:h_ar}. We have indeed found an induced geometry in the scalar part of the action dictated by the QM and background field, in a manner similar to that found for the Feynman propagator case. There are differences owing to the differing QM and diagonalization scheme however. Let us first comment that the two are the same in the low momenta limit. One may find (using QED units for comparison purposes) that $h_{ab}|_{|\bm{p}|\ll mc}=(q^2/4m^4c^4)F_{ai}F^{i}_{\;b}$. In this limit apart from the mass, we can see that the energy-momentum of the electromagnetic fields directly influences the induced geometry. The two induced geometries of either~\eqref{eq:h_f21} (2+1) or~\eqref{eq:h_ar}, however, differ as the spatial momentum increases.
The avenue to curved indices, and a curved space interpretation is the same as it was for the Feynman case in Sec.~\ref{sec:induced}. Namely one may do a change of coordinates using a similar tetrad basis, effectively producing a similar Lagrangian as given in~\eqref{eq:Leff_curved}. 
And the path integral measure changes dually as $\int^x_y\mathcal{D}x\mathcal{M}[x]\to\int^x_y\mathcal{D}x\sqrt{|g^{\rm ind}|}\mathcal{M}[x]$.

Finally, let us discuss the advanced propagator. The key difference mathematically from the retarded propagator is that the einbein is constrained to negative values, and hence after acquiring the $\delta[e-\dot{x}_0]$ function we are likewise restricted to negative $\dot{x}_0$, or $\theta(-z_0)$ paths. Next when performing the remaining momentum integrals the classical paths will follow $\hbar\bm{k}_{cl}=\mp\dot{\bm{x}}v_{F}^{-1}\Delta(\dot{x}_0^{2}-\dot{\bm{x}}^{2})^{-1/2}$, and hence the classical action $\pm v_{F}^{-1}\Delta\sqrt{\dot{x}_0^{2}-\dot{\bm{x}}^{2}}$. The fluctuation determinant is the same. Ultimately, one can find the same induced geometry. For completeness let us go ahead and write the full propagator
\begin{align}
    S^\textrm{adv}(x,y)&=-i\theta(-z_0)\sum_{r=\pm} P_{\bm{k}1}^r u_{\bm{k}1}\nonumber \\&
    \quad
    \times\int^x_y\!\mathcal{D}x\mathcal{M}[x]\,
    \bar{\mathcal{P}}\e^{\frac{i}{\hbar}\int^1_0d\tau\,\mathcal{L}_{\rm eff}^{(-r)}}u_{\bm{k}0}^\dagger\gamma_0\,,
    \label{eq:S_adv_final}
\end{align}
with anti-path ordering now in propertime dictated with $\bar{\mathcal{P}}$, and effective worldline action given by~\eqref{eq:Lr_eff} with $r\to-r$.

\subsection{(2+1)-dimensions}

Let us finally turn our attention to the (2+1)-dimensional case. The Hamiltonian,~\eqref{eq:H_2+1}, of this system is conventionally diagonalized in spherical coordinates in momentum space, however, for comparison let us maintain a similar eigendecomposition scheme as employed above. A quick note on conventions: in this section we define ${\bm \alpha}\equiv (\alpha^1,\alpha^2)$ over two dimensions and also for momentum such that ${\bm k}\equiv (k^1,k^2)$. Immediately then we can see that the similarity transform taking $u^\dagger_{\bm k}H_{\bm k}u_{\bm k}=E_{\bm k}\beta$ is simply given by~\eqref{eq:uk} and~\eqref{eq:ukd}, but in (2+1)-dimensions. Then we may directly find for the Berry phase and its correction that
\begin{equation}
    B^{i}=b^i\beta\,,\quad b^i=\frac{(\hbar v_{F})^{2}}{2E_{\bm{k}}(E_{\bm{k}}+\Delta)}\varepsilon^{ij}k_{j}\,,
\end{equation}
with ${\bm \delta B}$ given in the same form as~\eqref{eq:dB_ra}, but in (2+1)-dimensions. In polar coordinates one may find the usual Berry phase with angular dependence: ${\bm B}=(\hbar v_F)^2|{\bm k}|/[2E_{\bm k}(E_{\bm k}+\Delta)]\hat{{\bm \theta}}$. The curvature is given as
\begin{equation}
    S^{12}=\frac{(\hbar v_{F})^{2}\Delta}{2E_{\bm{k}}^3}\beta\,,
\end{equation}
the QM again takes the form~\eqref{eq:qgm3}, with spatial indices restricted to $i,j=1,2$. Projecting onto the energy sector $r=\pm1$ gives the phase-space Lagrangian~\eqref{eq:L_W_split} with $\mathbf B$ replaced by $r\mathbf b$. The kernels retain the form~\eqref{eq:W_H_split}, without path ordering. Similarly the path towards integrating out the momentum is almost the same as it was for the (3+1)-dimensional case. The replacements as dictated in~\eqref{eq:B_replace_ra} and~\eqref{eq:qm_replace_ra} are the same. However, afterwards when integrating out the remaining momenta d.o.f. we have a two-dimensional integral as represented in~\eqref{eq:3k_integrals}. Here the classical path is similar, however, the fluctuation determinant will instead become
\begin{equation}
    \mathcal{M}[x]=\det{}\Biggl[\frac{\Delta}{v_F}\frac{|\dot{x}_0 |}{\dot{x}_0^{2}-\dot{\bm{x}}^{2}}\Biggr]\,.
\end{equation}
The configuration-space Lagrangian retains the form~\eqref{eq:Lr_eff}, with $a,b=0,1,2$ and the branch-dependent Berry connection $\mathcal{B}_a^{(r)}=-rq\,b_iF^{i}{}_{a}/\hbar$ in place of $\mathcal{B}_a$. The retarded and advanced propagators follow from~\eqref{eq:S_ret_final} and~\eqref{eq:S_adv_final}, respectively, using the measure and omitting path and anti-path ordering.

\section{Magnetic null direction and spectral holonomy}
\label{sec:holonomy}

The induced metric has two linked consequences for propagation in electromagnetic backgrounds. 
\begin{enumerate}
    \item Its deformation has a magnetic null direction fixed by the local field and the particle four-velocity.
    \item Crossed fields also generate a mixed time--space component whose circulation lifts the degeneracy of counterpropagating modes.
\end{enumerate}
The first result is an exact tensor constraint whereas the second provides a global spectral probe of the same covariant geometry.

On the leading mass shell $p^2=M^2$, with $M=mc$, define $u^a = p_{\rm cl}^a / M$ so that $u^a u_a =1$. The quantum-metric tensor of the preceding configuration-space reduction~\eqref{eq:qgm2} then satisfies $\mathfrak{g}_{ab}u^b=0$ and the mass-shell pole is carried by $P^+ = \tfrac{1}{2}(1+\gamma^a u_a)$. With sign convention of~\eqref{eq:QGMreplace}, the induced metric
~\eqref{eq:induced_effective_metric} has deformation
\begin{equation}
    h_{\mu\nu}=-\left(\frac{q}{M}\right)^2F_\mu{}^a \mathfrak{g}_{ab}F^b{}_\nu ,
\end{equation}
indices in these contractions are raised with the background metric, and the complementary block being regular at $p=M$ produces no additional metric. The symmetric force product removes the Clifford commutator, so the $\mathcal{O}(\hbar^2 F^2)$ correction is proportional to the identity on the pole subspace and preserves its internal degeneracy. To make the factorization explicit, define $P_{ab}=(\eta_{ab}-u_au_b)/(2M)$. Then $\mathfrak g_{ab}=-P^c{}_aP_{cb}$, and the force contraction can be written as
\begin{equation}
    \mathcal Q_{a\mu} \coloneqq \frac{q}{M}P_a{}^bF_{b\mu},\qquad h_{\mu\nu}=-\mathcal Q^a{}_{\mu}\mathcal Q_{a\nu}.
    \label{eq:h_fact}
\end{equation}
Along a Lorentz-force trajectory, $dp_b=qF_{b\mu}dx^\mu$, and hence $\mathcal Q_{a\mu}dx^\mu= \tfrac{1}{M}P_a{}^bdp_b = \tfrac{dp_a}{2 M^2}$. 
Thus $\mathcal Q_{a\mu}dx^\mu$ is the transverse momentum change with the normalization fixed by the quantum metric.
Since $P_{ab}$ has rank three, then $ {\rm rank}\, h \leq3$ and $\det h=0$. 
This constraint applies to the deformation $h_{\mu\nu}$ while the complete metric remains nondegenerate when the perturbation is small in the reference orthonormal frame.

The direction annihilated by $h_{\mu\nu}$ can be identified directly from the electromagnetic field. Let us define 
\begin{equation}
    e^\mu=F^{\mu\nu}u_\nu,\qquad b^\mu={}^\star\!F^{\mu\nu}u_\nu
\end{equation}
are the electric and magnetic fields measured by an observer moving with four-velocity $u^\mu$ and ${}^{\star}\!F^{\mu\nu}=\frac{1}{2}\varepsilon^{\mu\nu\rho\sigma}F_{\rho\sigma}$ is the dual field. Both are orthogonal to $u^\mu$, and their spatial components in its rest frame are $\boldsymbol E_{\rm rest}$ and $\boldsymbol B_{\rm rest}$. Because,
\begin{equation}
    F_{a\nu}{}^\star\!F^{\nu\rho}=-\frac{1}{4}\delta_a{}^\rho F_{cd}{}^\star\!F^{cd}
\end{equation}
one has that $F_{a\nu}b^\nu$ is parallel to $u_a$. Since $\mathfrak{g}_{ab}u^b=0$, the deformation satisfies $h_{\mu\nu}b^\nu=0$. For $b^\mu\neq0$, the metric correction therefore vanishes along the magnetic field in the instantaneous rest frame. This direction is fixed by the local field and the particle velocity. The magnitude sampled by the trajectory is fixed by the rest-frame electric field. With $e_ae^a=-(E'_{\rm rest}/c)^2$ and $E_{\rm S}=M^2c/(|q|\hbar)$, one finds
\begin{equation}
    \hbar^2 h_{\mu\nu}u^\mu u^\nu=-\frac{\hbar^2q^2}{4M^4}e_ae^a=\frac{1}{4}\left(\frac{E'_{\rm rest}}{E_{\rm S}}\right)^2\equiv\frac{\chi^2}{4},
\end{equation}
where the dimensionless ratio $\chi$ controls the expansion. The perturbative metric applies for $\chi\ll1$ and becomes order unity at the pair-production scale. It is worth noting that smallness of the full deformation also requires control of its magnetic components; a purely magnetic field in the particle's rest frame has $\chi=0$ and can still produce a non-zero spatial deformation.

A gapped three-dimensional Dirac material realizes the corresponding Hamiltonian  $H=v_F\boldsymbol{\alpha}\cdot \boldsymbol{p}+\Delta\beta$, with $M=\Delta/v_F$ and $x^0=v_F t$. The field scales and expansion parameters are
\begin{equation}
    E_c =\frac{\Delta^2}{e\hbar v_F},\;\, B_c =\frac{E_c}{v_F},\;\, 
    \boldsymbol{\varepsilon}_E =\frac{\boldsymbol E}{E_c}, \;\,\boldsymbol{\varepsilon}_B =\frac{\boldsymbol B}{B_c}.
\end{equation}
Here $E_c$ is characteristic electric-field scale for Landau--Zener interband transitions. 
Near the band edge, $F_{0i}=E_i/v_F$ and $F_{ij}=-\epsilon_{ijk}B_k$ give
\begin{equation}\label{eq:met_deformation}
\begin{aligned}
    \hbar^2 h_{00}&=\frac{\varepsilon_E^2}{4},\\
    \hbar^2 h_{0i}&=-\frac{(\boldsymbol{\varepsilon}_E\!\times\!\boldsymbol{\varepsilon}_B)_i}{4},\\
    \hbar^2 h_{ij}&=\frac{\varepsilon_B^2\delta_{ij}-\varepsilon_{B,i}\varepsilon_{B,j}}{4}.
\end{aligned}
\end{equation}
Equivalently, $\hbar^2 h_{\mu\nu}dx^\mu dx^\nu=\frac{1}{4}\left|\boldsymbol{\varepsilon}_E dx^0+d\boldsymbol{x}\!\times\!\boldsymbol{\varepsilon}_B\right|^2$.
The spatial components obey $h_{ij}B_j=0$, while $h_{0i}$ is directed along $-\boldsymbol E\times\boldsymbol B$.
For comparison of counterpropagating trajectories, consider stationary fields and use the common band-edge metric~\eqref{eq:met_deformation}. Define $\mathcal A_i=g^{\rm ind}_{0i}/g^{\rm ind}_{00}$, and the optical path element
\begin{equation}
    dL = \sqrt{-\frac{1}{g^{\rm ind}_{00}}\bigg(g^{\rm ind}_{ij}-\frac{g^{\rm ind}_{0i}g^{\rm ind}_{0j}}{g^{\rm ind}_{00}}\bigg)dx^i dx^j}\,.
\end{equation}
For a particle with speed $v_{\rm loc}$ measured by stationary observers of this metric, the coordinate-time increment is 
\begin{equation}
    v_Fdt=-\mathcal{A}_i dx^i+\frac{v_F}{v_{\rm loc}}dL.
\end{equation}
The element $dL$ is unchanged under reversal of the other spatial displacement, and therefore cancels between two trajectories that follow the same closed path with equal local speed magnitudes at corresponding positions. However, the mixed term changes sign contributing twice to the difference between the two propagation times, given the resulting in the relative delay $\Delta t =t_+ - t_-$, 
\begin{equation}
\begin{aligned}
    \Delta t &=-\frac{2}{v_F}\oint_C\frac{g^{\rm ind}_{0i}}{g^{\rm ind}_{00}}dx^i\\ 
    &=\frac{1}{2v_F}\oint_C(\boldsymbol{\varepsilon}_E\!\times\!\boldsymbol{\varepsilon}_B)\cdot d\boldsymbol{\ell}+\mathcal{O}(\varepsilon^4).
\end{aligned}
\label{eq:crossed_field_holonomy}
\end{equation}
where $C$ is oriented along the + trajectory and the second line uses the retained band-edge metric.

Under a stationary change of time coordinate $x^0\mapsto x^0+f(\boldsymbol{x})$, the one-form transforms as $\mathcal A_i\mapsto \mathcal A_i-\partial_i f$. Its locally exact part can be removed, whereas a nonzero circulation cannot be eliminated by any single-valued $f$. The closed-loop integral is thus coordinate invariant resulting in a geometric contribution to propagation delay.

\subsection{Geometric splitting of counterpropagating modes}
Magnetic flux through an annulus produces an Aharanov--Bohm phase that changes the quantization condition for circulating modes and splits the energies of counterpropagating modes. The mixed components of the induced metric $g^{\mathrm{ind}}_{0i}$, couple the mode energy to its azimuthal momentum, so one can expect it modifies the splitting by odd-contributions in the circulation direction.

\begin{figure}
    \centering
    \includegraphics[width=0.7\linewidth]{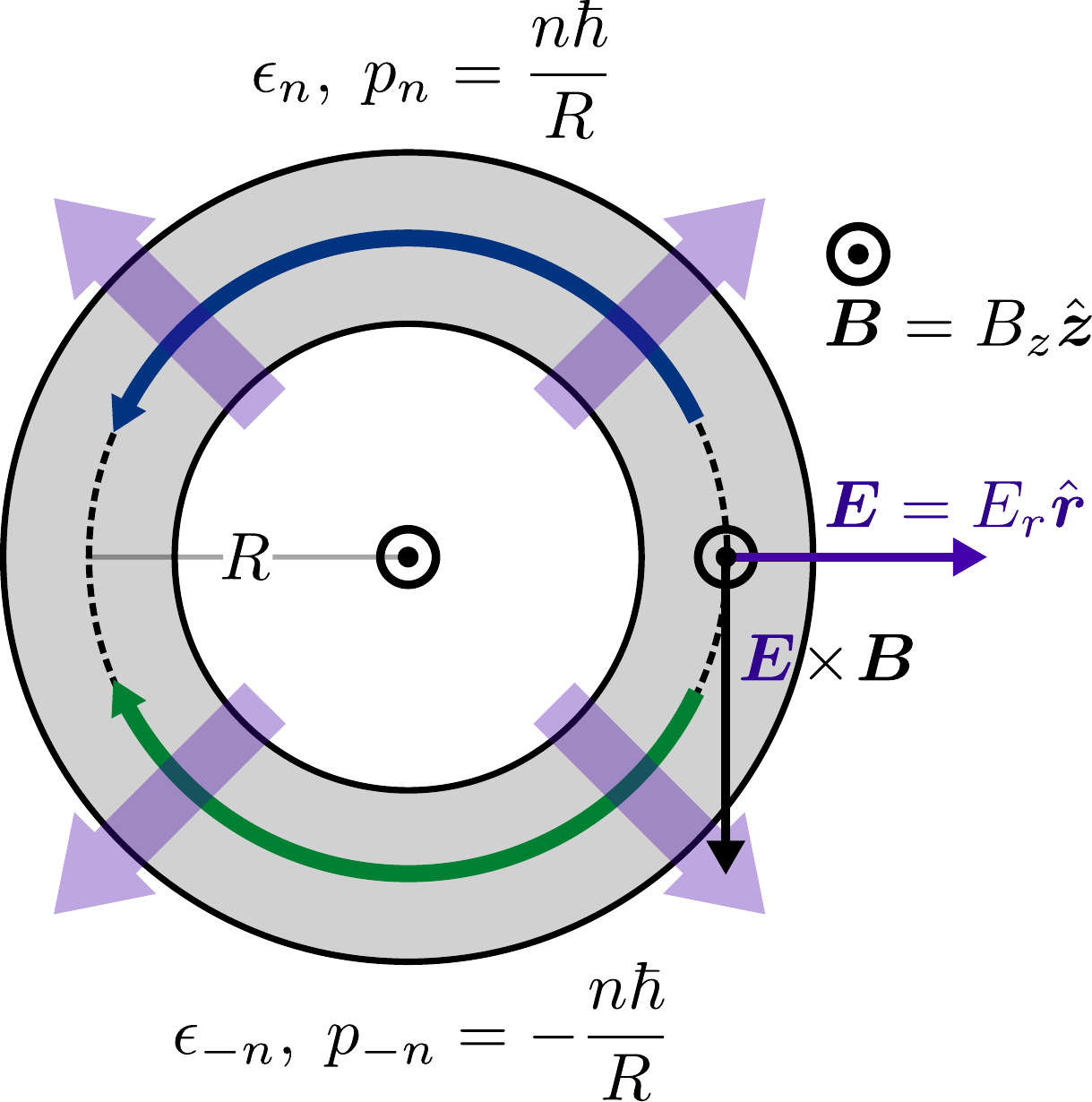}
    \caption{Counterpropagating modes in an annulus subject to radial electric and perpendicular magnetic, fields. The induced time--azimuth metric component contributes an energy splitting proportional to $E_r B_z$ between the $\pm$ modes.}
    \label{fig:mode_splitting}
\end{figure}

Consider a narrow annulus of radius $R$ with $\boldsymbol E=E_r\hat{\boldsymbol r}$ and $\boldsymbol B=B_z\hat{\boldsymbol z}$ as shown in~Fig.\ref{fig:mode_splitting}. Let $s=R\varphi$ and $a=\varepsilon_E\varepsilon_B/4$, with the signs of $E_r$ and $B_z$ included. Since $\boldsymbol{\varepsilon}_E\times\boldsymbol{\varepsilon}_B=-\varepsilon_E\varepsilon_B\hat{\boldsymbol\varphi}$, the relevant metric components are
\begin{equation}
    g_{\mu \nu}^{\rm ind} =
    \begin{pmatrix}
    1+\varepsilon_E^2 /4 & a \\
    a & -1+\varepsilon_B^2 /4
    \end{pmatrix}.
\end{equation} 
For a stationary mode, we can write its action as
\begin{equation}
    S_n=-\epsilon_nt+n\hbar\varphi=-\frac{\epsilon_n}{v_F}x^0+p_ns,
\end{equation}
where $p_n=n\hbar/R$ and $\epsilon_n$ is the quasiparticle energy and $p_n$ is the momentum along the annulus. The kinetic momentum is $\Pi_\mu=\partial_\mu S_n + qA_\mu$. At leading semiclassical order it satisfies 
\begin{equation}
    g_{\rm ind}^{\mu\nu}\Pi_\mu\Pi_\nu=M^2.
    \label{eq:annulus_HJ}
\end{equation}
Define $\mathcal{E}_n = \epsilon_n -q v_F A_0$, $\kappa_B = -q A_s$, and $\pi_n = p_n - \kappa_B$, with potentials evaluated on the annulus. Using the inverse band-edge metric through quadratic order in fields gives,
\begin{equation}
    0\simeq \Big(1-\frac{\varepsilon_E^2}{4}\Big)\frac{\mathcal E_n^2}{v_F^2}-\Big(1+\frac{\varepsilon_B^2}{4}\Big)\pi_n^2 -\frac{2a}{v_F}\mathcal E_n\pi_n-M^2 .
    \label{eq:annulus_hj_full}
\end{equation}
This expression includes the electrostatic shift, the ordinary Aharanov--Bohm response, the pure electric and magnetic metric corrections, and the mixed metric component. We now expand the canonical energy as
\begin{equation}
    \epsilon_n \simeq \epsilon_n^{{}_{(0)}}+\delta\epsilon_n^{{}_{(E)}}+\delta\epsilon_n^{{}_{(B)}}+\delta\epsilon_n^{{}_{(EB)}}.
\label{eq:annulus_energy_expansion}
\end{equation}
The zero-field energy obeys $\epsilon_n^{{}_{(0)}}=v_F\sqrt{p_n^2+M^2}$. At first order in $E_r$, the electrostatic shift is identified as $\delta\epsilon_{n,1}^{{}_{(E)}}=qv_FA_0$, while at second order the temporal metric component produces
\begin{equation}
    \delta\epsilon_n^{{}_{(E)}}=qv_F A_0 + \frac{\varepsilon_E^2}{8}\epsilon_n^{{}_{(0)}} , 
\end{equation}
both terms are even in $n$ and cancel in the energy difference between opposite circulation directions.

For the magnetic contribution, write $\delta\epsilon_n^{{}_{(B)}}=\delta\epsilon_{n,1}^{{}_{(B)}}+\delta\epsilon_{n,2}^{{}_{(B)}}$, with subscripts denoting the order in $B_z$. The linear equation is
\begin{equation}
    0=\frac{2\epsilon_n^{{}_{(0)}}}{v_F^2}\delta\epsilon_{n,1}^{{}_{(B)}}+2p_n\kappa_B,
\end{equation}
giving the ordinary Aharanov--Bohm energy shift $\delta\epsilon_{n,1}^{{}_{(B)}}=-v_F^2p_n \kappa_B / \epsilon_n^{{}_{(0)}}$. At quadratic order,
\begin{equation}
    0=\frac{2\epsilon_n^{{}_{(0)}}}{v_F^2}\delta\epsilon_{n,2}^{{}_{(B)}}+\frac{(\delta\epsilon_{n,1}^{{}_{(B)}})^2}{v_F^2}-\kappa_B^2 - \frac{\varepsilon_B^2}{4}p_n^2,
\end{equation}
which gives
\begin{equation}
\delta\epsilon_{n,2}^{{}_{(B)}}=\frac{v_F^2}{2\epsilon_n^{{}_{(0)}}}\left[\kappa_B^2-\frac{(\delta\epsilon_{n,1}^{{}_{(B)}})^2}{v_F^2}+\frac{\varepsilon_B^2}{4}p_n^2\right].
\end{equation}
The term linear in $\kappa_B$ is $n$-odd and produces the conventional Aharonov--Bohm splitting. The quadratic magnetic terms are even in $n$ and cancel in the pair separation. Last, we collect the terms proportional to $E_rB_z$ in~\eqref{eq:annulus_hj_full} resulting in
\begin{equation}
    0=\frac{2\epsilon_n^{{}_{(0)}}}{v_F^2}\delta\epsilon_n^{{}_{(EB)}}+\frac{2}{v_F^2}\left(\delta\epsilon_{n,1}^{{}_{(E)}}-qv_FA_0\right)\delta\epsilon_{n,1}^{{}_{(B)}}-\frac{2a\epsilon_n^{{}_{(0)}}p_n}{v_F}.
\end{equation}
The first-order electrostatic shift cancels inside the parentheses, leaving
\begin{equation}
    \frac{2\epsilon_n^{{}_{(0)}}}{v_F^2}\delta\epsilon_n^{{}_{(EB)}}-\frac{2a\epsilon_n^{{}_{(0)}}p_n}{v_F}=0 .
\end{equation}
Hence the mixed metric contribution is
\begin{equation}
    \delta\epsilon_n^{{}_{(EB)}}=\frac{n\hbar v_F}{4R}\varepsilon_E\varepsilon_B.
\end{equation}
The mode with opposite angular momentum is obtained by $n\to-n$, giving
\begin{equation}
    \delta\epsilon_n^{{}_{(EB)}}-\delta\epsilon_{-n}^{{}_{(EB)}}=\frac{n\hbar v_F}{2R}\varepsilon_E\varepsilon_B.
\end{equation}
The splitting reverses under either $E_r\to-E_r$ or $B_z\to-B_z$ and grows linearly with $n$. Taking the part odd in both fields removes the electrostatic shift and the ordinary Aharonov--Bohm contribution.

The annulus also connects this spectral result to the propagation delay derived above. For counterpropagating trajectories compared at equal local speeds, this circulation produces a propagation delay~\eqref{eq:crossed_field_holonomy} of $\Delta t\simeq-4\pi Ra/v_F$. For stationary modes, the same circulation enters the periodic quantization condition and results in the energy splitting proportional to $E_rB_z$ derived above.

\section{Conclusions}
\label{sec:conclusions}

We have studied quantum geometry, viz. the Berry phase and QM, appearing in a Lorentz covariant construction of the fermionic path integral. The Berry phase appears in a quantum mechanical representation in momentum space, and therefore an intuitive extension that incorporates time on the same footing as space is provided by the phase space worldline formalism, in which for the Feynman representation the effective `Hamiltonian' is $\slashed{p}$. We furthermore studied several different systems of interest to both condensed matter and QED applications, including in addition to the Feynman propagator the advanced/retarded propagators, and for each we examined both the (2+1) and (3+1)-dimensional representations. Applying the adiabatic expansion to the worldline action yields the Berry connection at $\mathcal{O}(\hbar)$ and the QM at $\mathcal{O}(\hbar^2)$. We found the phase and QM differed upon diagonalization for the Feynman and advanced/retarded propagators due to whether the effective Hamiltonian was diagonalized before or after the on-shell criteria, $p^2=m^2c^2$, was enforced.

A key finding of our analysis is that upon integrating out the momentum d.o.f. from the path integral with accompanying QM is the appearance of an induced Lorentzian geometry in configuration space. Since the QM term in the effective worldline Lagrangian for both the Feynman and advanced/retarded cases is proportional to $\dot{p}\cdot \mathfrak{g} \cdot \dot{p}$ for QM, $\mathfrak{g}_{ab}$, it is intuitive and we find that the $\dot{p}_a$ becomes field dependent according to the Lorentz force e.o.m. so that the induced geometry goes as $\dot{x}\cdot F\cdot \mathfrak{g}\cdot F\cdot \dot{x}$. The primary difference between the Feynman and advanced/retarded cases occurs from diagonalizing either according to respectively the mass-shell condition or to energy. We have last shown phenomenological characteristics of the induced geometry that affect the magnetic null direction and spectral holonomy. 

We have exploited a direct connection between the QM and a unique induced geometry in the \textit{fermionic} sector; however in theory the techniques employed here can be extended to a variety of systems in which a Berry phase appears. One need only to analyze a first quantized construction of said systems under the semi-classical limit projecting as well the QM. One key example of interest lies in QCD where color d.o.f. can be diagonalized. In the worldline formalism the Wilson line of the quark can be Abelian projected wherein $\mathcal{O}(\hbar^2)$ corrections are conjectured, as reported here, to give rise to a QM--directly in configuration space. This may have implications for the low-energy and non-perturbative sector of QCD.

\begin{acknowledgments}	
We would like to thank Yuki Amari, Chihiro Sasaki and participants of the ``Avenues of Quantum Field Theory in Curved Spacetime 2026" workshop for fruitful discussions. PC would also like to acknowledge support from the WPI program ``Sustainability with Knotted Chiral Meta Matter (WPI-SKCM${}^2$)" at Hiroshima University, and the Research Start-up Support Fund of WPI-SKCM${}^2$.
\end{acknowledgments}

\appendix

\section{Bloch metric and covariant pullback}
\label{app:bloch-metric}

For the planar Dirac Hamiltonian~\eqref{eq:H_2+1}, with $\Delta>0$, the band projectors are
\begin{equation}
    P_\pm(\mathbf{k})=\tfrac12[1\pm\hat{\mathbf d}\cdot\boldsymbol{\sigma}],
\end{equation}
with $\hat{\bf d}=\tfrac{\hbar v_F}{E_k}(k_x,k_y,\bar{\Delta})$ where $\bar \Delta \coloneqq \Delta/ (\hbar v_F)$. Using $\operatorname{tr}(\sigma^\alpha\sigma^\beta)=2\delta^{\alpha\beta}$, both bands have the positive quantum metric
\begin{equation}
    g^{{\rm FS}, k}_{ij}=\frac{1}{2}{\rm tr}\{(\partial_{k_i}P_{\pm})(\partial_{k_j}P_{\pm})\}=\frac14\partial_{k_i}\hat{\mathbf d}\cdot\partial_{k_j}\hat{\mathbf d},
\end{equation}
where $i,j=1,2$. Evaluating the derivatives using $\partial_{k_i}E_k=(\hbar v_F)^2k_i/E_k$ gives $g^{{\rm FS},k}_{ij}=-\mathfrak g^{H,k}_{ij}$, with the signed Hamiltonian tensor given in~\eqref{eq:qgm3}.

To compare with the covariant metric, set $p^a=(p^0,\pi_1,\pi_2,0)$, with $M=\Delta/v_F$, $\boldsymbol{\pi}=\hbar\mathbf{k}$, and $p^0=E_k/v_F$. On the mass shell $p=M$, the condition $p_a dp^a=0$ reduces~\eqref{eq:qgm2} to $ds_p^2=-dp_a dp^a/(4M^2)$. Substituting $dp^0=\pi_i d\pi_i/p^0$ gives
\begin{equation}
    ds_p^2=\frac{1}{4M^2}\left(\delta_{ij}-\frac{\pi_i\pi_j}{(p^0)^2}\right)d\pi_i d\pi_j.
\end{equation}
This recovers the covariant pullback in~\eqref{eq:metric_comparison}, with
\begin{equation}
    g^{\mathrm{FS},\pi}_{ij}=\frac{1}{\hbar^2} g^{\mathrm{FS},k}_{ij}=\frac{\Delta^2}{E_k^2}\mathfrak g^{F,\Sigma}_{ij}.
\end{equation}
The two positive tensors therefore coincide at the band edge, where $g^{\mathrm{FS},k}_{ij}(0)=\ell^2\delta_{ij}/4$ and $\ell=\hbar v_F/\Delta$ is the interband length.

\bibliography{references}
\bibliographystyle{apsrev4-2}
\end{document}